\documentclass[11pt,a4paper]{article}

\usepackage{vmargin}
\setmarginsrb{2.6cm}{1.7cm}{2.6cm}{1.7cm}{1cm}{1cm}{1cm}{1.6cm}

\usepackage{amsmath}
\usepackage{amsthm,amssymb,epsfig,latexsym}
\newtheorem{prop}{Proposition}[section]
\newtheorem{lemma}{Lemma}[section]
\newtheorem{cor}{Corollary}[section]
\newtheorem{Def}{Definition}[section]
{\theoremstyle{remark}
\newtheorem{rem}{Remark}[section]}

\def\qed{\hfill\nobreak\hbox{$\square$}\par\medbreak}
 \makeatletter
 \@addtoreset{equation}{section}
 \makeatother

\newcommand{\be}[1]{\begin{equation}\label{#1}}
\newcommand{\ba}[1]{\begin{multline}\label{#1}}
\newcommand{\ee}{\end{equation}}
\newcommand{\ea}{\end{eqnarray}}

\newcommand{\num}{\\\rule{0pt}{20pt}}

\newcommand{\dis}{\displaystyle}
\newcommand{\eq}[1]{(\ref{#1})}
\newcommand{\tr}{\mathop{\rm tr}}

\newcommand{\f}[2]{{\ensuremath{%
    \mathchoice%
    {\dfrac{#1}{#2}}
    {\dfrac{#1}{#2}}
    {\frac{#1}{#2}}
    {\frac{#1}{#2}}
}}}
\newcommand{\tf}[2]{\ensuremath{#1/#2}}
\newcommand{\pa}[1]{\ensuremath{\left(#1\right)}}
\newcommand{\paa}[1]{\ensuremath{\left\{#1\right\}}}
\newcommand{\pac}[1]{\ensuremath{\left[#1\right]}}
\newcommand{\paf}[2]{\ensuremath{\left(\f{#1}{#2}\right)}}
\def\Ga{\Gamma}

\def\la{\lambda}

\def\sg{\sigma}

\def\Om{\Omega}
\def\om{\omega}
\def\vp{\varphi}

\newcommand{\mc}[1]{\ensuremath{\mathcal{#1}}}

\newcommand{\wt}[1]{\ensuremath{\widetilde{#1}}}
\newcommand{\wh}[1]{\ensuremath{\widehat{#1}}}

\newcommand{\Int}[2]{\ensuremath{\int\limits_{#1}^{#2}}}

\newcommand{\pl}[2]{\ensuremath{\prod\limits_{#1}^{#2}}}

\newcommand{\R}{\ensuremath{\mathbb{R}}}

\newcommand{\s}[1]{\ensuremath{\sinh\pa{#1}}}

\newcommand{\ket}[1]{|\,#1\,\rangle}
\def\tr{\operatorname{tr}}

\newcommand{\e}[1]{\ensuremath{\mathrm{#1}}}

\begin{document}

\begin{flushright}
LPENSL-TH-03/10\\
DESY 11-037
\end{flushright}
\par \vskip .1in \noindent

\vspace{14pt}

\begin{center}
\begin{LARGE}
{\bf  Thermodynamic limit of particle-hole form factors\\ in the
massless XXZ Heisenberg chain}
\end{LARGE}

\vspace{30pt}

\begin{large}

{\bf N.~Kitanine}\footnote[1]{IMB, UMR 5584 du CNRS, Universit\'e
de Bourgogne, France, Nicolai.Kitanine@u-bourgogne.fr},~~
{\bf K.~K.~Kozlowski}\footnote[2]{DESY, Hamburg, Deutschland,
 karol.kajetan.kozlowski@desy.de},~~
{\bf J.~M.~Maillet}\footnote[3]{ Laboratoire de Physique, UMR 5672
du CNRS, ENS Lyon,  France,
 maillet@ens-lyon.fr},\\
{\bf N.~A.~Slavnov}\footnote[4]{ Steklov Mathematical Institute,
Moscow, Russia, nslavnov@mi.ras.ru},~~
{\bf V.~Terras}\footnote[5]{ Laboratoire de Physique, UMR 5672 du
CNRS, ENS Lyon,  France, veronique.terras@ens-lyon.fr}
\par

\end{large}

\vspace{40pt}

\centerline{\bf Abstract} \vspace{1cm}
\parbox{12cm}{\small  We study the thermodynamic limit of the particle-hole form factors of the XXZ
Heisenberg chain in the massless regime. We show that, in this
limit, such form factors decrease as an explicitly computed power-law in the system-size.
Moreover, the corresponding amplitudes  can be obtained as a product of a
``smooth" and a ``discrete" part: the former depends continuously on
the rapidities of the particles and holes, whereas the latter has an
additional explicit dependence on the set of integer numbers that label each excited state in the associated
logarithmic Bethe equations. We also show that special
form factors corresponding to zero-energy excitations lying on the Fermi surface decrease as a power-law in the system size with the same critical
exponents as %the ones appearing
in the long-distance asymptotic behavior of the related
two-point correlation functions. The methods we develop in this article are rather general and can be applied  to other
massless integrable models associated to the six-vertex $R$-matrix and having determinant representations for their form factors. }
\end{center}

\vspace{40pt}

\section{Introduction\label{INT}}

This article is a continuation of our study of form  factors in
massless quantum integrable models. We have recently
\cite{KitKMST09c} investigated the thermodynamic limit of a special,
Umklapp-type, form factor of the $XXZ$ Heisenberg spin chain
\cite{Hei28,Bet31,Hul38,Orb58,Wal59,CloG66,YangY66,LieM66} starting
from its determinant representation  \cite{KitMT99} in the algebraic
Bethe Ansatz framework \cite{FadST79,FadLH96,BogIK93L}. We now apply
this method to general, particle-hole type, form factors. The main
goal of our analysis is to pave a way for the calculation of the
asymptotic behavior of correlation functions in massless integrable
models through their form factor expansion. Our results should also
be useful in the numerical computation of dynamical structure
factors through their (exact) form factor expressions starting from
finite size systems
\cite{CauM05,CauHM05,PerSCHMWA06,PerSCHMWA07,CauCS07} as they allow
a precise {\it a priori}   control of the dependence of each form
factor in terms of the system size.

For definiteness, we chose to focus on a particular model, the XXZ
spin-$1/2$ Heisenberg chain \cite{Hei28} in the massless regime and
in an external magnetic field $h>0$ for which determinant
representations of general form factors were obtained in
\cite{KitMT99}. However, up to minor modifications specific to the
choice of the model, our method and results apply as well to other
massless, algebraic Bethe Ansatz-solvable models  with known
determinant representations of their form factors like the
non-linear Schr\"odinger model \cite{KojKS1997,KorS1999} or the
higher spin XXX \cite{CasM2007} and XXZ \cite{DegM2009} chains.

The Hamiltonian of the XXZ chain is given by,
 \begin{equation}\label{0-HamXXZ}
 H=\sum_{k=1}^{M}\left(
 \sigma^x_{k}\sigma^x_{k+1}+\sigma^y_{k}\sigma^y_{k+1}
 +\Delta(\sigma^z_{k}\sigma^z_{k+1}-1)\right)-\frac{h}{2}\sum_{k=1}^{M}\sigma^z_{k} \; ,
 \end{equation}
where $\sigma^{x,y,z}_{k}$ are the spin operators (Pauli matrices) acting on the $k^{\e{th}}$ site
of the chain.
The main purpose of this article is to study the asymptotic behavior
of the form factors in the thermodynamic limit $M\to\infty$ of the chain and in the massless regime  $|\Delta|<1$.
We will take this limit starting from a chain of finite even size and subject to periodic boundary conditions.

Considering the finite chain allows us to define rigorously the form factors of
local spin operators as the normalized matrix elements
 \begin{equation}\label{def-FF-stand}
 {\cal F}^{(s)}_{\psi\,\psi'}(m)=\frac{\langle\,\psi \, |\, \sigma_m^s\, |\, \psi' \rangle}
                                                           {\|\psi\|\cdot \|\psi'\|} \;,\qquad
 s=x,y,z,
 \end{equation}
where $\ket{\psi}$ and $|\, \psi'\rangle$ are two eigenstates of the
Hamiltonian \eqref{0-HamXXZ}. Such matrix elements can be computed
in a systematic way by using  the solution to the quantum inverse
scattering problem \cite{KitMT99,MaiT00} together with the explicit
formulae for scalar products \cite{Sla89,Sla95,KitMT99} in the
algebraic Bethe Ansatz framework
\cite{FadST79,FadLH96,BogIK93L,Sla07}.

Using the closure relation, one can express any zero temperature two-point correlation function of spin operators
as a sum over the form factors of the corresponding local operators,
\begin{equation}\label{Cor-fun}
 \frac{\langle\psi_g \, |\, \sigma_{m}^s\, \sigma_{m'}^{s'}\, | \, \psi_g\rangle}
 {\langle\psi_g\,|\,\psi_g\rangle}=
 \sum_{|\,\psi'\rangle} {\cal F}^{(s)}_{\psi_g\,\psi'}(m)\; {\cal F}^{(s')}_{\psi'\,\psi_g}(m').
 \end{equation}
Here $\ket{\psi_g}$ denotes the ground state of the Hamiltonian \eqref{0-HamXXZ},
and the summation is taken with respect to all the eigenstates $|\,\psi^{\prime}\rangle$
of \eqref{0-HamXXZ}.

One possible way of dealing with \eqref{Cor-fun} was proposed in
\cite{KitMST05a,KitMST05b}, where the sum over the form factors for
the finite chain was recast into a multiple contour integral, the
so-called master equation. This method then leads to various
multiple integral representations for the correlation functions in
the thermodynamic limit \cite{KitMST05a,KitMST05b,KitMST05c} that
are in fact non-trivial summations of previously obtained multiple
integral representations for the correlation functions
\cite{JimMMN92,JimM96,JimM95L,KitMT00,KitMST02a,KitMST02b}.
Moreover, in \cite{KitKMST09b} we have shown how to derive
long-distance asymptotic behavior of certain two-point correlation
functions using such  master equation representation. It agrees with
predictions arising from the Luttinger-liquid  and Conformal Field
Theories approaches
\cite{LutP75,Hal80,Hal81a,Hal81b,Aff85,BloCN86,Car84,Car86}.  It is
the very remarkable structure of the  results obtained in
\cite{KitKMST09b} (the amplitudes of the power law decrease of the
long distance behavior of the two-point  correlation function are
related to the modulus squared of some properly normalized form
factors) that strongly suggested that another possible way to
analyze the asymptotic behavior of the two-point correlation
functions would be to take the thermodynamic limit directly in
\eqref{Cor-fun}. It seems indeed quite natural that, in this limit,
the main contribution to the sum \eqref{Cor-fun} should stem from
excitations above the ground state having a finite energy. Such
excited states can be characterized in terms of particles, holes
and/or string configurations. In this framework, the sum over the
complete set of states in \eqref{Cor-fun} becomes a sum over all the
possible particle-hole-string type excitations above the ground
state which, after some possible regularizations, should be
replaceable by integrals in the thermodynamic limit. Another
important motivation to learn how to deal directly with the form
factor expansion of the correlation functions concerns the time
dependent case, as in that situation the asymptotic analysis through
the dynamical master equation \cite{KitMST05b} poses yet unsolved
problems. The first step to carry out this program is therefore to
determine the leading asymptotic behavior of the form factors of the
model in the thermodynamic limit. It is the purpose of the present
article to solve this problem for the so-called particle/hole type
form factors (corresponding to diffusion states). The case of states
associated with complex solutions of the Bethe equations such as
string type solutions (corresponding to bound states) requires the
use of additional techniques and will be considered in a separate
publication.

In \cite{KitKMST09c}, we have  shown that, in the $M\to \infty$
limit, a special Umklapp-type form factor of the $\sg^z$ operator
decreases as some negative power of the system size $M$. The
striking feature of this analysis is that this power coincides with
one of the critical exponents appearing in the long-distance
asymptotic behavior of the zero temperature two-point correlation
function $\langle\sigma_{1}^z\sigma_{m+1}^{z}\rangle$. In the
present article, we will show that general form factors exhibit
similar properties: they all decay as some negative power of the
size $M$; moreover, for special types of form factors corresponding,
in the thermodynamic limit,  to particle-hole excitations lying on
the Fermi surface, these exponents coincide exactly with the
critical exponents describing  the power-law decay of the
long-distance asymptotic behavior of the corresponding two-point
correlation functions.

More precisely, in the thermodynamic limit,  the product of two form
factors appearing in \eqref{Cor-fun} can be presented in the
following form
 \begin{equation}\label{ff-struct}
 {\cal F}^{(s)}_{\psi_g\,\psi'}(m') \cdot {\cal F}^{(s')}_{\psi'\,\psi_g}(m)
 =M^{-\theta_{ss'}}\, e^{i{\cal P}_{ex}(m-m')}\,{\cal S}_{ss'}\,{\cal D}_{ss'}.
 \end{equation}
All the dependence on the lattice distance $m-m'$ is contained in an
obvious phase factor\footnote[1]{%
Some form factors (see e.g. \eqref{ff-fin}) may also have an
additional factor $(-1)^{m-m'}$, which can be removed by the
re-definition of the Hamiltonian \eqref{0-HamXXZ} }. There ${\cal
P}_{ex}$ is the excitation momentum. The finite part of the form
factors product can be separated into two parts ${\cal S}_{ss'}$ and
${\cal D}_{ss'}$. We call them smooth and discrete parts
respectively. We show that the smooth part ${\cal S}_{ss'}$ depends
continuously on the rapidities of the particles and holes. On the
contrary, it is only when the particles (holes) rapidities are
separated from the Fermi boundary that the discrete part ${\cal
D}_{ss'}$ does also depend smoothly on these quantities. As soon as
the latter approach the Fermi surface, the discrete structure of the
form factors reveals itself in ${\cal D}_{ss'}$. This means that a
microscopic (of order $1/M$) deviation of a particle (hole)'s
rapidity leads to a macroscopic change of ${\cal D}_{ss'}$. We
believe that one of the consequences of such a structure is the
particular role played by the Fermi boundary in bosonization
techniques.

The article is organized as follows. In section~\ref{DoES} we give a
description of the low-lying excited states of the model. Following
the ideas of \cite{KitKMST09c}, we consider excitations above the
ground state of the twisted transfer matrix. This enables us to
formulate precisely our results in section~\ref{SFinRes}, i.e. the
structure of the large size asymptotic behavior of general
particle-hole form factors in the XXZ spin-$\tf{1}{2}$ chain. The
remaining part of the article is devoted to the proof of this
result. In section~\ref{FFaSP} we show how form factors can be
reduced to scalar products of a special type that we can represent in
terms of determinants for chains of finite length $M$. We then
calculate the thermodynamic limit  of the form factors of the
$\sigma^z$ operator in section~\ref{ThDL-Sz}. In particular, we show
that if the rapidities  of the particles and holes collapse on the
Fermi surface then the form factors have a discrete structure.
In section~\ref{WWS+1/2} we obtain analogous results for the form factors of the $\sigma^\pm$ operators\footnote[2]{%
Recall that $\sigma^\pm=\frac12(\sigma^x\pm i\sigma^y)$}. We would
like to stress that several methods and results used in the present
article are borrowed from our previous work \cite{KitKMST09c}. We
conclude by discussing the possibility of applying our results to
the calculation of the long-distance asymptotic behavior of the
two-point functions at zero temperature. Several auxiliary Lemmas
and  proofs of technical character are presented in the two
appendices.

%%%%%%%%%%%%%%%%%%%%%%%%%%%%%%%%%%%%%%%%%%%%%%%%%%%%%%%%%%%%%%%%%%%%%%
%%%%%%%%%%%%%%%%%%%%%%%%%%%%%%%%%%%%%%%%%%%%%%%%%%%%%%%%%%%%%%%%%%%%%%
%%%%%%%%%%%%%%%%%%%%%%%%%%%%%%%%%%%%%%%%%%%%%%%%%%%%%%%%%%%%%%%%%%%%%%
%%%%%%%%%%%%%%%%%%%%%%%%%%%%%%%%%%%%%%%%%%%%%%%%%%%%%%%%%%%%%%%%%%%%%%
%%%%%%%%%%%%%%%%%%%%%%%%%%%%%%%%%%%%%%%%%%%%%%%%%%%%%%%%%%%%%%%%%%%%%%
%%%%%%%%%%%%%%%%%%%%%%%%%%%%%%%%%%%%%%%%%%%%%%%%%%%%%%%%%%%%%%%%%%%%%%

\section{Space of states of the model\label{DoES}}

Since our ultimate goal is to analyze the two-point correlation
functions at zero temperature through \eqref{Cor-fun}, the form
factors we consider here correspond to matrix elements of a local
operator between the ground state and an excited eigenstate of the
Hamiltonian \eqref{0-HamXXZ}. In this section we recall how such
states can be described in the large $M$ limit.

In the algebraic Bethe Ansatz framework
\cite{FadST79,FadLH96,BogIK93L}, the space of states of the model
(in finite volume) is  constructed by means of the Yang-Baxter
algebra realized by the operator entries of the monodromy matrix. In
the XXZ case, and more generally for the class of models associated
with the six-vertex R-matrix, such monodromy matrices take the form
 \begin{equation}\label{Mon-Mat}
 T(\lambda)=\begin{pmatrix} A(\lambda)& B(\lambda)\\
                                                   C(\lambda)& D(\lambda)\end{pmatrix},
 \end{equation}
where $A$, $B$, $C$, $D$ are quantum operators depending on some
spectral parameter $\la$. It satisfies Yang-Baxter commutation
relations given by the six-vertex R-matrix. In this framework, the
eigenstates $|\psi\rangle$ of the Hamiltonian \eqref{0-HamXXZ}
coincide with the ones of the transfer matrix ${\cal T}(\lambda)=\tr
T(\lambda)=A(\la)+D(\la) $, and can be parameterized as
$|\psi\rangle=|\psi(\lambda_1,\dots,\lambda_N)\rangle$,
$N=0,1,\dots,M/2$, in terms of a set $\{\la_j\}_{j=1}^N$ of
solutions to the logarithmic Bethe equations
 \begin{equation}\label{BE}
 Mp_0(\lambda_j)-\sum_{k=1}^N\vartheta(\lambda_j-\lambda_k)=2\pi n_j,
 \qquad j=1,\dots,N.
 \end{equation}
Here the functions $p_0(\lambda)$ and $\vartheta(\lambda)$ are the bare
momentum and phase,
 \begin{equation}\label{1-p0}
 p_0(\lambda)=i\log\left(\frac{\sinh(\textstyle{\frac{i\zeta}2}+\lambda)}
 {\sinh(\textstyle{\frac{i\zeta}2}-\lambda)}\right),\qquad
 \vartheta(\lambda)=i\log\left(\frac{\sinh(i\zeta+\lambda)}
 {\sinh(i\zeta-\lambda)}\right),
 \end{equation}
where $0<\zeta<\pi$ and $\cos\zeta =\Delta$.
The numbers $n_j$, $-M/2<n_j\le M/2$, are integers (for  $N$ odd)
or half-integers (for  $N$ even).

\subsection{Thermodynamic limit of the ground state\label{ThLGS}}

The Bethe roots $\la_j$, $j=1,\ldots,N$, describing the ground
state correspond to the solution of the system of logarithmic Bethe
Ansatz equations \eqref{BE} with the special choice $n_j=j-(N+1)/2$,
where $N$ is the number of down spins
in the ground state\footnote{%
The number $N$ depends on the magnetization of the ground state,
which is fixed by the overall magnetic field $h$. Hereafter, we call
$N$ sector the subspace of the space of states having $N$ spins
down.} \cite{Hul38,LieSM61,LieM66,YangY66}. Thus, for given $M$ and
$h$, these parameters $\la_j$ are fixed quantities. In order to
describe how they behave in the thermodynamic limit $(N,M\to\infty$
such that $N/M$ tends to some fixed average density $D$), we
introduce the ground state counting function,
 \begin{equation}\label{BE-cf}
 \wh{\xi}(\omega)
    =\frac1{2\pi}p_0(\omega)-\frac1{2\pi M}\sum_{k=1}^N\vartheta(\omega-\lambda_k)
    +\frac{N+1}{2M},
 \end{equation}
which is build so that  $\wh{\xi}(\lambda_j)=j/M$ for  $j=
1,\dots,N$. This function defines the discrete density of the minimal energy state in the $N$ sector
by $\wh{\rho}(\omega)=\wh{\xi}\,'(\omega)$. It is possible to argue
that, in the thermodynamic limit, the parameters $\lambda_j$ fill
densely a finite interval $[-q,q]$ of the real axis (the Fermi
zone), and that the discrete density goes to a smooth function
$\rho(\omega)$ that solves the following integral equation
 \begin{equation}\label{Int-eq-Dden}
 \rho(\lambda)+\frac1{2\pi}\int\limits_{-q}^qK(\lambda-\mu)\,\rho(\mu)\,d\mu=\frac1{2\pi}p'_0(\lambda),
 \quad  \e{with} \quad K(\lambda)=\vartheta'(\lambda).
 \end{equation}
The value of the endpoint $q$ of the Fermi zone is fixed by
requiring that $\int_{-q}^{q} \rho\pa{\la}=D$. Note that, since
$\lim_{M\to \infty} \wh{\xi}\pa{\la_1}=0$, the thermodynamic limit
$\xi(\lambda)$ of the counting function is the antiderivative of
$\rho(\lambda)$ that vanishes at $-q$. Therefore $\xi(\lambda)$ can
be expressed in terms of the dressed momentum $p(\lambda)$, the Fermi momentum being defined by  $k_F = p(q)$
 \begin{equation}\label{0-Dmom}
 p(\lambda)=2\pi\int\limits_0^\lambda \rho(\mu)\,d\mu,\qquad
 \xi(\lambda)=[p(\lambda)+p(q)]/2\pi, \qquad \pi D=p(q)\equiv k_F .
 \end{equation}

Several other properties of the ground state can be described through the solutions of other linear integral equations.
We introduce two such functions that will play an important role in our analysis,
namely the dressed charge $Z(\lambda)$ and the dressed phase
$\phi(\lambda,\nu)$, which satisfy
 \begin{equation}\label{Int-eq-DC}
 Z(\lambda)+\frac1{2\pi}\int\limits_{-q}^qK(\lambda-\mu)\, Z(\mu)\,d\mu=1  ,
 \end{equation}
and
 \begin{equation}\label{Int-eq-DF}
\phi(\lambda,\nu)+\frac1{2\pi}\int\limits_{-q}^qK(\lambda-\mu)\, \phi(\mu,\nu)\,d\mu=\frac1{2\pi}\vartheta(\lambda-\nu)  .
 \end{equation}
In fact, these two functions are not independent. Using that
$K(\lambda)=\vartheta'(\lambda)$, one easily obtains that
 \begin{equation}\label{Z-Th}
 Z(\lambda)=1+\phi(\lambda,q)-\phi(\lambda,-q)  .
 \end{equation}
Another non-trivial relationship between $Z$ and $\phi$ has been established in
\cite{KorS98,Sla98}:
 \begin{equation}\label{prop-phase}
 1+\phi(q,q)-\phi(-q,q)=Z^{-1}(q) .
 \end{equation}

\subsection{Thermodynamic limit of the excited states\label{ThLES}}

To describe the other eigenstates of the transfer matrix
$\mathcal{T}(\la)$, it is convenient for further purposes to
consider instead the twisted transfer matrix ${\cal
T}_\kappa(\lambda)=A(\la)+\kappa D(\la)$, where $\kappa$ is some
complex (twist) parameter. Hereby the excited states of the
Hamiltonian \eqref{0-HamXXZ} are obtained as the $\kappa\to 1$ limit
of the eigenstates of the twisted transfer matrix  ${\cal
T}_\kappa(\lambda)$ \cite{KitMST05a,KitKMST09c}.

In complete analogy with the standard algebraic Bethe Ansatz
considerations, the eigenstates $|\psi_\kappa(\{\mu\})\rangle$ of
${\cal T}_\kappa(\lambda)$ can be parameterized by sets of solutions
of the twisted Bethe equations
 \begin{equation}\label{TBE-lj}
 Mp_0(\mu_{\ell_j})-\sum_{k=1}^{N_\kappa}\vartheta(\mu_{\ell_j}-\mu_{\ell_k})=2\pi \left(\ell_j-
 \frac{{N_\kappa}+1}2\right)+2\pi\alpha,
 \qquad j=1,\dots,{N_\kappa}.
 \end{equation}
Such states $|\psi_\kappa(\{\mu\})\rangle$ depend on $N_\kappa$
parameters $\paa{\mu_{\ell_k}}_{k=1}^{N_{\kappa}}$, where $N_\kappa$
is the number of spins down and we have set $\kappa=e^{2\pi
i\alpha}$. In this article we consider $\alpha$ to be a real number.
This restriction $\Im(\alpha)=0$ is not crucial, but convenient,
since then $\langle \psi_\kappa(\{\mu\})|=(-1)^{N_\kappa}
|\psi_\kappa(\{\mu\})\rangle^\dagger$.
\begin{rem}
One could of course from the very beginning set $\alpha=0$  and deal
directly with the standard excited states of the XXZ Hamiltonian
\eqref{0-HamXXZ}. However, we chose to introduce this extra
parameter and keep it throughout our computations since, as we will
show later on, an important class of form factors can be obtained by
a mere shift of $\alpha$ by some integer value.
\end{rem}

In the following, we will consider two cases
of interest: $N_\kappa=N$ for form factors of $\sigma^z$, and
$N_\kappa=N+1$ for the $\sigma^+$ form factors\footnote[1]{%
We do not consider the case $N_\kappa=N-1$ corresponding to form
factors of $\sigma^-$, since the last ones can be obtained from the
form factors of $\sigma^+$.}. It follows from \cite{YangY66}  that
the set of  real roots of the twisted Bethe equations is completely
defined by the choice of the set of integers $\ell_j$ in the
\textit{r.h.s.} of \eqref{TBE-lj}. Therefore, we have labeled the
roots of \eqref{TBE-lj} by subscripts $\ell_j$.

\begin{Def}\label{Def-aGS}
The state parameterized by the real  solutions to \eqref{TBE-lj}
with $\ell_j=j$, $j=1,\dots,N_{\kappa}$, is called $\alpha$-twisted
ground state in the $N_\kappa$ sector.
\end{Def}

In the thermodynamic limit, excitations corresponding to real set of
solutions to the twisted Bethe equations above this $\alpha$-twisted
ground state can be described in terms of particles and holes. Note
that the states associated to  complex roots of the Bethe equations
(like string solutions) are not considered in the present article.
They correspond in particular to form factors of  bound states
\cite{Orb58,Wal59,CloG66,BabdVV83} that will be dealt with in a
separate publication. All this means that one deals with solutions
to the twisted Bethe equations where most of the $\ell_{j}$ coincide
with their value for the ground state: $\ell_j=j$ except for  $n$
integers for which $\ell_j\ne j$, with $n$ remaining finite in the
$N_\kappa\to\infty$ limit. To be more precise, we fix $2n$ distinct
integers $h_1,\dots,h_n$ and $p_1,\dots,p_n$ such that $h_k\in
\{1,\dots,{N_\kappa}\}$  and $p_k\notin \{1,\dots,{N_\kappa}\}$. The
integers $h_k$ represent "holes" in respect to the distributions of
integers for the ground state in the $N_\kappa$ sector, whereas the
integers $p_k$ represent "particles". In other words, the excited
state is described by a set of integers
$\paa{\ell_{j}}_{j=1}^{N_{\kappa}}$ such that $\ell_j=j$ for $j\ne
h_1,\dots,h_n$, and $\ell_{h_k}=p_k$.

The terminology of particles and holes takes a clearer
interpretation  in terms of the counting function
$\wh{\xi}_{\kappa}$ associated with the ${N_\kappa}$-excited state:
 \begin{equation}\label{TBE-cf}
\wh{\xi}_\kappa(\omega)
    =\frac1{2\pi}p_0(\omega)-\frac1{2\pi M}\sum_{k=1}^{{N_\kappa}}\vartheta(\omega-\mu_{\ell_k})
    +\frac{{N_\kappa}+1}{2M}-\frac{\alpha}{M} \; .
   \end{equation}
One can argue that the counting function is monotonously increasing
on the real axis.  Thus it is possible to define unambiguously a set
of real parameters $\mu_j$ as the unique solutions to
$\wh{\xi}_{\kappa}\pa{\mu_j}=\tf{j}{M}$. One can identify, among
this set of parameters, the $N_{\kappa}$ solutions of the twisted
Bethe equations with integers $\ell_j$ as being $\mu_{\ell_j}$ as it
should be. In this picture, an excited state no-longer corresponds
to the set of $N_{\kappa}$ consecutive solutions $\mu_j$ to the
equation $\wh{\xi}_{\kappa}\pa{\mu_j}=\tf{j}{M}$, but is rather
obtained by removing the solutions
$\wh{\xi}_{\kappa}\pa{\mu_{h_a}}=\tf{h_a}{M}$ and replacing them by
the solutions $\wh{\xi}_{\kappa}\pa{\mu_{p_a}}=\tf{p_a}{M}$. It is
in this respect that $\mu_{h_a}$ stands for the rapidities of the
holes and that $\mu_{p_a}$ stands for those of the particles. From
now on, we agree upon $\mu_j$ being the solution to
$\wh{\xi}_{\kappa}\pa{\mu_j}=\tf{j}{M}$ with $\wh{\xi}_{\kappa}$
being given by \eqref{TBE-cf}.

Similarly as for the ground state in the $N$ sector, one can also
define the discrete density of the excited state by
$\wh\rho_\kappa(\omega)=\wh{\xi}\,'_\kappa(\omega)$. It is easy to
see that the thermodynamic limits of this new counting function and
its associated density coincide with the limits defined above in
\eqref{Int-eq-Dden}, \eqref{0-Dmom}.

An excited state is most conveniently characterized in terms of the
shift function $\wh F(\omega)=\wh F(\omega|\{\mu_p\}|\{\mu_h\})$
defined by
 \begin{equation}\label{SP-shiftF-mod}
 \wh
 F(\omega)=M\bigl(\wh\xi(\omega)-\wh\xi_\kappa(\omega)\bigr).
 \end{equation}
The shift function describes the spacing between the root $\la_j$ for the  ground state in the $N$ sector
and the parameters $\mu_j$ defined by $\wh{\xi}_{\kappa}\pa{\mu_j}=\tf{j}{M}$:
 \begin{equation}\label{shift-TD}
 \mu_j-\lambda_j=\frac{F(\lambda_j)}{\rho(\lambda_j)M}+O(M^{-2}),
 \end{equation}
where $F(\lambda)$ is the thermodynamic limit of the shift function.
Recall that we consider the excited states in the
$N_\kappa$ sector with $N_\kappa=N$ (for form factors of
$\sigma^z$) and $N_\kappa=N+1$ (for form factors of $\sigma^+$).
Respectively  we should distinguish between two shift functions
$F^{(z)}(\lambda)$ and $F^{(+)}(\lambda)$ corresponding to these two
cases. Generically the shift function $F(\lambda)$ satisfies the
integral equation
 \begin{equation}\label{Int-eq-F}
 F(\lambda)+\int\limits_{-q}^qK(\lambda-\mu)F(\mu)\,\frac{d\mu}{2\pi}
 =\alpha+\frac{\delta N}{2}\left[1-\frac{\vartheta(\lambda- q) }{\pi}\right]+
 \frac1{2\pi}\sum_{k=1}^{n}\big[ \vartheta(\lambda- \mu_{p_k})-\vartheta(\lambda-\mu_{h_k}) \big]
.
 \end{equation}
%
%
%\frac1{2\pi}\sum_{k=1}^{n}
where $\delta N=N-N_{\kappa}$. Using \eqref{Int-eq-DC} and
\eqref{Int-eq-DF} we conclude that
 \begin{equation}\label{F-solution}
 F^{(z)}(\lambda)=\alpha Z(\lambda)+
 \sum_{k=1}^{n}\phi(\lambda,\mu_{p_k})-\sum_{k=1}^n\phi(\lambda,\mu_{h_k}) \;.
 \end{equation}
and
 \begin{equation}\label{F-solution+}
 F^{(+)}(\lambda)=\left(\alpha-\frac{1}2\right) Z(\lambda)+
 \sum_{k=1}^{n}\phi(\lambda,\mu_{p_k})-\sum_{k=1}^n\phi(\lambda,\mu_{h_k}) + \phi(\la,q)\;.
 \end{equation}
%

%%%%%%%%%%%%%%%%%%%%%%%%%%%%%%%%%%%%%%%%%%%%%%%%%%%%%%%%%%%%%%%%%%%%%%
%%%%%%%%%%%%%%%%%%%%%%%%%%%%%%%%%%%%%%%%%%%%%%%%%%%%%%%%%%%%%%%%%%%%%%
%%%%%%%%%%%%%%%%%%%%%%%%%%%%%%%%%%%%%%%%%%%%%%%%%%%%%%%%%%%%%%%%%%%%%%
%%%%%%%%%%%%%%%%%%%%%%%%%%%%%%%%%%%%%%%%%%%%%%%%%%%%%%%%%%%%%%%%%%%%%%
%%%%%%%%%%%%%%%%%%%%%%%%%%%%%%%%%%%%%%%%%%%%%%%%%%%%%%%%%%%%%%%%%%%%%%
%%%%%%%%%%%%%%%%%%%%%%%%%%%%%%%%%%%%%%%%%%%%%%%%%%%%%%%%%%%%%%%%%%%%%%

\section{The main results \label{SFinRes}}

We now formulate a precise statement of the results  obtained in
this article concerning the thermodynamic behavior of form factors.
In fact, for later use,  we will give this thermodynamic behavior
directly for the products of form factors appearing in the spectral
expansion of the correlation function \eqref{Cor-fun}.  However, it
will be made clear that our method allows also to obtain the
behavior of each individual form factor. The proof of the results we
present below  will be given in the next sections.

We consider the form factors \eqref{def-FF-stand} of  the local spin
operators $\sigma_m^s$, $s=+,-,z$, between the ground state
$\ket{\psi_g}$ and some (different) excited state $\ket{\psi'}$ of
the Hamiltonian\footnote[1]{%
The case when $\ket{\psi'}=\ket{\psi_g}$ is trivial: ${\cal
F}^{(\pm)}_{\psi_g\,\psi_g}(m)=0$,  ${\cal
F}^{(z)}_{\psi_g\,\psi_g}(m)=2D-1$.}. Here and in the following, we
set  $\ket{\psi_g}\equiv \ket{\psi(\{\la\} ) }$,
$\{\la_1,\ldots,\la_N\}$ being the solution of  the Bethe equations
\eqref{BE} describing the ground state of the Hamiltonian, and
$\ket{\psi'}=\lim_{\kappa\to 1}\ket{\psi_\kappa(\{\mu\})}$ for
$\{\mu_{\ell_1},\ldots,\mu_{\ell_{N_\kappa} }\}$ being  a solution
of the $\kappa$-twisted Bethe equations \eqref{TBE-lj} with $n$
particles and $n$ holes. We will be specially interested in the
limiting case where, in the thermodynamic limit, all rapidities
$\{\mu_{p_j}\}_{j=1}^n$ and $\{\mu_{h_j}\}_{j=1}^n$ of particles and
holes condensate on the Fermi boundaries. Such excited states have
a zero excitation energy in the thermodynamic limit; we expect that they will produce the main
contribution to the asymptotic behavior of correlation functions
(this fact is already apparent in the predictions based on the
Luttinger liquid theory and Conformal Field Theory for the long-distance asymptotic behavior of
two-point functions \cite{LutP75,Hal80,Hal81a,Hal81b,Aff85,BloCN86,Car84,Car86}). We will
also consider the opposite limiting case, namely when all
particle/hole rapidities remain at finite distance from the Fermi
boundaries.

In the thermodynamic limit, up to uniformly $M^{-1}\cdot \log M$ corrections, the products of such form factors behave as
 \begin{equation}\label{ff-fin}
 \begin{array}{l}
 {\dis {\cal F}_{\psi_g\,\psi'}^{(z)}(m') \cdot {\cal F}_{\psi'\,\psi_g}^{(z)}(m)
  \sim \delta_{N ,N_\kappa}\; M^{-\theta_{zz}} \, e^{i{\cal P}_{ex}(m-m')} \;
    \Bigl.\partial^2_\alpha \, {\cal S}_{zz} \, {\cal D}_{zz}\Bigr|_{\alpha=0}   \, ,}\num
 {\dis {\cal F}^{(+)}_{\psi_g\,\psi'}(m') \cdot {\cal F}^{(-)}_{\psi'\,\psi_g}(m)
  \sim \delta_{N+1 , N_\kappa}\; M^{-\theta_{+-}}\, (-1)^{m-m'} e^{i{\cal P}_{ex}(m-m')} \;
  \Bigl.{\cal S}_{+-} \, {\cal D}_{+-}\Bigr|_{\alpha=0} \, .}
 \end{array}
 \end{equation}

Let us explain these formulas. These products decrease as some
negative power of the size $M$ of the system, the exponents
$\theta_{zz}$, $\theta_{+-}$ being specified below. Due to the
translation invariance of the model, all the dependence on the
lattice spacing $m-m'$ is absorbed into a phase factor, with
 \begin{equation}\label{mom-ex}
    {\cal P}_{ex}= \lim_{M,N \rightarrow +\infty} \Bigl\{\sum_{j=1}^{N_\kappa} p_0(\mu_{\ell_j})-\sum_{j=1}^N p_0(\la_j)
    \Bigr\}
                          = 2\pi\alpha D+ \sum_{j=1}^{n} \big[ p(\mu_{p_j})-p(\mu_{h_j}) \big]
 \end{equation}
being the momentum of the excited state relative to the ground
state. We have split the constant in front of the  $M$ and $(m-m')$
dependence  in two parts, ${\cal S}_{ss'}$ and ${\cal D}_{ss'}$,
that we call smooth and discrete parts respectively. The reason for
this denomination is that the smooth part ${\cal S}_{ss'}$ depends
continuously on the rapidities $\mu_{p_j}$ and $\mu_{h_j}$ of the
particles and holes, whereas the discrete part ${\cal D}_{ss'}$ also
depends on the set of integers appearing in the logarithmic Bethe
Ansatz equations \eqref{TBE-lj} for the excited state. Such a
discrete structure can be fairly well approximated by a smooth
function of the rapidities $\mu_{p_j}$ and $\mu_{h_j}$ as long as
they are located at a finite distance from the Fermi boundaries.
However, as soon as the rapidities of the  particles or holes
approach the Fermi surface, the discrete structure of the form
factors can no longer be neglected: a microscopic (of order $1/M$)
deviation of a particle (or hole) rapidity leads to a macroscopic
change in ${\cal D}_{ss'}$.

%%%%%%%
\subsection{Smooth parts}

The smooth parts ${\cal S}_{zz}$ and ${\cal
S}_{+-}$ are decomposed into
 \begin{equation}\label{smooth-parts}
 {\cal S}_{zz}=\frac{2}{\pi^2}\sin^2\pa{\frac{{\cal P}_{ex}}2} \cdot
 {\cal A}^{(z)}_n \cdot e^{ C^{(z)}_n } ,  \qquad \quad
  {\cal S}_{+-}= {\cal A}^{(+)}_n \cdot  e^{C_n^{(+)}}.
 \end{equation}
Here $C_n^{(z)}$, ${\cal A}_n^{(z)}$ are functionals acting on the
shift function $F^{\pa{z}}(\la)$ given in \eqref{F-solution}, while
$C_n^{(+)}$, ${\cal A}_n^{(+)}$ are functionals acting on the shift
function $F^{\pa{+}}(\la)$ given in \eqref{F-solution+}. This action
is such that the result  depends smoothly on the particle/hole
rapidities $\{\mu_p\}$ and $\{\mu_h\}$ of the state
$\ket{\psi_\kappa(\{\mu\})}$.

Before giving  explicit expressions for the quantities in
\eqref{smooth-parts} we introduce the $i\pi$-periodic Cauchy
transform of the shift functions ${F}^{(z/+)}(\lambda)$ on $[-q,q]$,
 \begin{equation}\label{tFbis}
 \mathbf{F}^{(z/+)}(w)=\frac1{2\pi i}\int\limits_{-q}^qF^{(z/+)}(\lambda)\coth(\lambda-w)\,d\lambda.
 \end{equation}
Then the coefficient $ C_{n}^{(z)}=C_{n}^{(z)}(\{\mu_p\},\{\mu_h\})$
has the form
 \begin{multline}\label{Cnz}
  C_{n}^{(z)}[F^{(z)}]
  =C_0[F^{(z)}]+2\pi i \sum_{j=1}^n\bigl(\mathbf{F}^{(z)}(\mu_{h_j}-i\zeta)+\mathbf{F}^{(z)}(\mu_{h_j}+i\zeta)
 -\mathbf{F}^{(z)}(\mu_{p_j}-i\zeta)-\mathbf{F}^{(z)}(\mu_{p_j}+i\zeta)\bigr)\\
 +\sum_{j,k=1}^n\log\frac{\sinh(\mu_{h_j}-\mu_{p_k}-i\zeta)\, \sinh(\mu_{p_k}-\mu_{h_j}-i\zeta)}
 {\sinh(\mu_{p_j}-\mu_{p_k}-i\zeta) \, \sinh(\mu_{h_j}-\mu_{h_k}-i\zeta)},
 \end{multline}
where the functional $C_0[F]$ reads
\begin{equation}\label{C0}
  C_0[F]=
 -\int\limits_{-q}^{q}\frac{F (\lambda) \, F (\mu)}{\sinh^2(\lambda-\mu-i\zeta)}\,
      d\lambda \, d\mu.
 \end{equation}

The coefficient $C_{n}^{(+)}=C_{n}^{(+)}(\{\mu_p\},\{\mu_h\})$ has
similar representation, but it contains several additional terms
 \begin{multline}\label{Cn+}
  C_{n}^{(+)}[F^{(+)}]=C_{n}^{(z)}[F^{(+)}]
   -2\pi i \bigl(\mathbf{F}^{(+)}(q+i\zeta)+\mathbf{F}^{(+)}(q-i\zeta)\big)\\
  +\sum_{j=1}^n\log\frac{\sinh(\mu_{h_j}-q-i\zeta)\,\sinh(\mu_{h_j}-q+i\zeta)}
                                        {\sinh(\mu_{p_j}-q-i\zeta)\,\sinh(\mu_{p_j}-q+i\zeta)}.
\end{multline}

The functional ${\cal A}_n^{(z)}={\cal
A}_n^{(z)}(\{\mu_p\},\{\mu_h\})$ has the form
 \begin{multline}\label{Anz}
  {\cal A}_n^{(z)}[F^{(z)}]
    =\left(\frac{\sin\pi\alpha }{\sin\pi F^{(z)}(-q)}\right)^2
 \left|e^{-2\pi i\mathbf{F}^{(z)}(-q+i\zeta)}\prod_{k=1}^n\frac{\sinh(q+\mu_{h_k}+i\zeta)}
 {\sinh(q+\mu_{p_k}+i\zeta)}\right|^2\\
   \times  \left|\frac{\det\big[I+\frac1{2\pi i}U^{(z)}(w,w')\big] }
    {\det \big[ I+\frac1{2\pi}K \big] }\right|^2.
  \end{multline}
The equation \eqref{Anz} contains a ratio of Fredholm determinants.
The integral operator $I+\frac{1}{2\pi} K$ acts on the interval
$[-q,q]$, whereas the integral operators $I+\frac{1}{2\pi i}
U^{(z)}$ acts  on a counterclockwise oriented contour $\Gamma_q$
surrounding $[-q,q]$. This contour is such that it contains all the
ground state roots and no other singularity of the kernel. The
integral kernel is defined by
 \begin{equation}\label{U-qz}
 U^{(z)}(w,w')=-\Phi(w) e^{2\pi i\bigl(\mathbf{F}^{(z)}(w)-\mathbf{F}^{(z)}(w+i\zeta)\bigr)} \frac{K_\kappa(w-w')-K_\kappa(-q-w')}
 {1-e^{2\pi iF^{(z)}(w)}} \;,
 \end{equation}
 with
 \begin{equation}\label{2-Kk}
  K_\kappa(w)=\coth(w+i\zeta)-\kappa\coth(w-i\zeta),
  \qquad \Phi(w)=\prod_{k=1}^n
 \frac{\sinh(w-\mu_{p_k})\sinh(w-\mu_{h_k}+i\zeta)}
 {\sinh(w-\mu_{h_k})\sinh(w-\mu_{p_k}+i\zeta)}.
 \end{equation}

The coefficient ${\cal A}_n^{(+)}={\cal
A}_n^{(+)}(\{\mu_p\},\{\mu_h\})$ also is proportional to a ratio of
Fredholm determinants
 \begin{equation}\label{An+}
{\cal A}_n^{(+)}[F^{(+)}]
 =\frac{\sin\zeta}{2\pi\kappa}
 \left| \frac{e^{-2\pi i\mathbf{F}^{(+)}(\frac{i\zeta}2)} }{\sinh(q-\frac{i\zeta}{2})}
 \prod\limits_{k=1}^n
 \frac{\sinh(\mu_{h_k}-\frac{i\zeta}2)}{\sinh(\mu_{p_k}-\frac{i\zeta}2)}
 \cdot \frac{\det\big[I+\frac1{2\pi i}U^{(+)}(w,w')\big] }
    {\det\big[I+\frac1{2\pi}K\big]}\right|^2 \!\! .
 \end{equation}
Similarly to the integral operator defined in \eqref{U-qz}, the
operator $I+\frac{1}{2\pi i} U^{(+)}$ acts on the contour $\Gamma_q$
surrounding $[-q,q]$, with a kernel
 \begin{equation}\label{kernel-3}
 U^{(+)}\pac{F}(w,w')= \Phi(w)\frac{\sinh(w-q)}{\sinh(w-q+i\zeta)}
               e^{2\pi i\bigl(\mathbf{F}^{(+)}(w)-\mathbf{F}^{(+)}(w+i\zeta)\bigr)}
\frac{{\cal K}_\kappa(w,w')}{1-e^{2\pi i
 F^{(+)}(w)}},
 \end{equation}
where
 \begin{equation}\label{K-2}
 {\cal K}_\kappa(w,w')=\frac{\sinh(w'+\frac{3i\zeta}2)}
 {\sinh(w'-\frac{i\zeta}2)\sinh(w-w'-i\zeta)}-
 \frac{\kappa\sinh(w'-\frac{3i\zeta}2)}
 {\sinh(w'+\frac{i\zeta}2)\sinh(w-w'+i\zeta)},
 \end{equation}
and $\Phi(w)$ is given by \eqref{2-Kk}.

%%%%%%
\subsection{Discrete parts and exponents}

The values of the discrete parts  ${\cal D}_{zz}$ and ${\cal
D}_{+-}$, as well as of the exponents $\theta_{zz}$ and
$\theta_{+-}$, depend on whether particles and holes are on the
Fermi surface or separated from it. We specify their expressions in
the two particular cases we mentioned above: when all particle/hole
rapidities remain at finite distance from the Fermi boundaries, or
when all particle/hole rapidities collapse on the Fermi boundaries. Other intermediate configurations can be obtained along similar techniques.

\subsubsection{Particles and holes away from the Fermi boundaries}

In the first case, the exponents in \eqref{ff-fin} giving the
algebraic decay of the form factor with the system size read:
 \begin{equation}\label{powers-1}
 \begin{array}{l}
 {\dis \theta_{zz}=2n+\left(F_+^{(z)}\right)^2+\left(F_-^{(z)}\right)^2,}\num
 {\dis \theta_{+-}=2n+\left(F^{(+)}_++1\right)^2+\left(F^{(+)}_-\right)^2,}
 \end{array}
 \end{equation}
%
%
%
%Above one should understand
where $F^{(z/+)}_{\pm}=F^{(z/+)}(\pm q)$ are given in terms of the shift functions defined in \eqref{F-solution} and \eqref{F-solution+}. The discrete part ${\cal
D}_{zz}$ is given by
 \begin{multline}\label{lim-Dz}
 {\cal D}_{zz} (\{\mu_p\},\{\mu_h\})=\left[\det_n\frac1{\sinh(\mu_{p_j}
 -\mu_{h_k})}\right]^2 \cdot
\mc{D}^{\pa{z}}[F^{\pa{z}}]
\prod_{k=1}^n\frac{\sin^2(\pi F^{(z)}(\mu_{h_k}))}{\pi^2\, \rho(\mu_{h_k})\, \rho(\mu_{p_k})} \\
 \times \prod_{k=1}^n
 \exp\left\{2\int\limits_{-q}^q\hspace{-4.2mm}\diagup\hspace{.5mm}F^{(z)}(\lambda)\bigl[\coth(\lambda-\mu_{p_k})-\coth(\lambda-\mu_{h_k})\bigr]
 \,d\lambda\right\}.
\end{multline}
Here   a  functional $\mc{D}^{\pa{z}}[F]$  has the following form:
\begin{equation}
\mc{D}^{\pa{z}}\pac{F}= \frac{G^2 (1-F_- ) \, G^2 (1+F_+)\, (2\pi)^{F_- - F_+ }}
 {[\rho(q)\, \sinh(2q)]^{\left(F_-\right)^2+\left(F_+\right)^2}}
 e^{C_1[F]}   \; .
\label{fonctionnelle Dz}
\end{equation}
where $G(z)$ is the Barnes function satisfying
$G(z+1)=\Gamma(z)G(z)$, and we have set
 \begin{multline}\label{7-C1}
 C_1[F]=\int\limits_{-q}^{q}
 \frac{{F}'(\lambda)F(\mu) -F(\lambda)\, {F}'(\mu)}{2\tanh(\lambda-\mu)}\,
 d\lambda \, d\mu\\
 +F_+\hspace{-1mm}\int\limits_{-q}^{q}\frac{F_+ - F(\lambda)}{\tanh(q-\lambda)}\,d\lambda
 +F_-\hspace{-1mm}\int\limits_{-q}^{q}\frac{F_- - F(\lambda)}{\tanh(q+\lambda)}\,d\lambda \ .
 \end{multline}

The discrete part ${\cal D}_{+-}$ has the form
 \begin{multline}\label{D-separ}
   {\cal D}_{+-}(\{\mu_p\},\{\mu_h\})
   =\left[\det_n\frac1{\sinh(\mu_{p_j}
 -\mu_{h_k})}\right]^2 \cdot
\mc{D}^{\pa{+}}[F^{\pa{+}}]
 \prod_{k=1}^n\frac{\sin^2(\pi F^{(+)}(\mu_{h_k}))}{\pi^2\, \rho(\mu_{h_k})\, \rho(\mu_{p_k})}
\\
 \times \prod_{k=1}^n \left[\frac{\sinh(\mu_{p_k}-q)}{\sinh(\mu_{h_k}-q)}\right]^2 \exp\left\{ 2\,\!\int\limits_{-q}^q \hspace{-4.2mm}\diagup\hspace{.5mm}F^{(+)}
 (\lambda)\bigl[\coth(\lambda-\mu_{p_k})-\coth(\lambda-\mu_{h_k})\bigr]
 \,d\lambda\right\},
  \end{multline}

where the functional $\mc{D}^{\pa{+}}[F]$ reads
\begin{equation}
\mc{D}^{\pa{+}}\pac{F} =
\mc{D}^{\pa{z}}\pac{F}\cdot\frac{\sinh(2q)\Gamma^2 (1+F_+)}
 { [\rho(q)\, \sinh(2q)]^{1+2F_+} }
% \frac{G^2 (1-F_- ) \, G^2 (2+F_+)\, (2\pi)^{F_- - F_+}}
% {[\rho(q)\, \sinh(2q)]^{\left(F_-\right)^2+\left(1+F_+\right)^2}}
 %
\exp\paa{2\int\limits_{-q}^{q}\frac{F_{+}-F(\lambda)}{\tanh(q-\lambda)}\,d\lambda
}\;.
\label{fonctionnelle D+}
\end{equation}

\subsubsection{Particles and holes on the Fermi boundaries}

\begin{Def}
Let an  excited state contain $n$ particles and $n$ holes. It is
called critical excited state, if all the rapidities
$\mu_p,\mu_h=\pm q$ in the thermodynamic limit. We also say that
this excited state belongs to the $\mathbf{P}_{r}$ class if it
contains $n^{\pm}_p$ particles, resp. $n^{\pm}_h$ holes, with
rapidities equal to $\pm q$ such that
 \begin{equation}\label{Nak-mom}
 n^+_p-n^+_h=n^-_h-n^-_p=r, \qquad r\in\mathbb{Z}.
 \end{equation}
The corresponding form factors are called \underline{critical form
factors} of the $\mathbf{P}_{r}$ class.
\end{Def}

The rapidities of the particles and holes in an excited state
belonging to the $\mathbf{P}_{r}$ class are all located in a close
neighborhood of $\pm q$. As a consequence, it is useful to
re-parameterize the  integers describing the position of particles
and holes according to
 \begin{equation}\label{spec-p}
 \begin{array}{ll}
 p_j=p_j^++N_\kappa,&\mbox{if}\quad \mu_{p_j}=q \, ,\\
 p_j=1-p_j^-,&\mbox{if}\quad \mu_{p_j}=-q \, ,\\
 h_j=N_\kappa+1-h_j^+,&\mbox{if}\quad \mu_{h_j}=q \, ,\\
 h_j=h_j^-,&\mbox{if}\quad \mu_{h_j}=-q \, .
 \end{array}
 \end{equation}
All the integers  $\{p^\pm\}$ and $\{h^\pm\}$ introduced above are positive and vary in a range such that
 \begin{equation}\label{domain}
 \lim_{N\to\infty}\frac{\sum p_j^\pm}N=\lim_{N\to\infty}\frac{\sum h_j^\pm}N=0 \,
 ,
 \end{equation}
which means that $\mu_{p}$ and $\mu_{h}$ indeed collapse to the Fermi
boundary in the thermodynamic limit.

For $\sigma^z$ and $\sigma^\pm$ form factors belonging to the
$\mathbf{P}_{r}$ class, the discrete parts are given by
 \begin{multline}\label{D-fin-z}
   {\cal D}_{zz}  = \mc{D}^{\pa{z}}[F^{\pa{z}}_{r}]
            \f{G^{2}(1+F^{\pa{z}}_+) G^{2}(1-F^{\pa{z}}_-) }
                        {G^{2}(1+F^{\pa{z}}_{r,+}) G^{2}(1-F^{\pa{z}}_{r,-})   }
 \left(\frac{\sin(\pi F^{(z)}_{r,+})}{\pi}\right)^{2n^+_h}
 \left(\frac{\sin(\pi F^{(z)}_{r,-})}{\pi}\right)^{2n^-_h}\\
 \times
 R_{n_p^+,n_h^+}(\{p^+\},\{h^+\}|F^{(z)}_+) \; R_{n_p^-,n_h^-}(\{p^-\},\{h^-\}|-F^{(z)}_-)\; ,
  \end{multline}
and
 \begin{multline}\label{D-fin+-}
 {\cal D}_{+-}   =\mc{D}^{\pa{+}}[F^{\pa{+}}_{r}]
         \f{G^{2}(2+F^{\pa{+}}_+) G^{2}(1-F^{\pa{+}}_-) }
                        {G^{2}(2+F^{\pa{+}}_{r,+}) G^{2}(1-F^{\pa{+}}_{r,-})   }
  \left(\frac{\sin(\pi F^{(+)}_{r,+})}{\pi}\right)^{2n_h^+}
  \left(\frac{\sin(\pi F^{(+)}_{r,-})}{\pi}\right)^{2n_h^-} \\
  \times
  R_{n_p^+,n_h^+}(\{p^+\},\{h^+\}|1+F^{(+)}_+) \; R_{n_p^-,n_h^-}(\{p^-\},\{h^-\}|-F^{(+)}_-) \; .
  \end{multline}
Here we agree upon  $F^{(z/+)}_{r}(\la)=F^{(z/+)}(\la)+r$ and
$F^{(z/+)}_{r,\pm}=F^{(z/+)}_r(\pm q)$. The functionals
$\mc{D}^{\pa{z}}$ and $\mc{D}^{\pa{+}}$ are given respectively in
\eqref{fonctionnelle Dz}  and \eqref{fonctionnelle D+}. The
coefficient $R_{n,m}(\{p\},\{h\}|F)$ is defined as
 \begin{equation}\label{def-R}
 R_{n,m}(\{p\},\{h\}|F)= \frac{\prod\limits_{j>k}^n(p_j-p_k)^2\prod\limits_{j>k}^m(h_j-h_k)^2}
 {\prod\limits_{j=1}^n\prod\limits_{k=1}^m(p_j+h_k-1)^2} \;
 \Gamma^2 \left( \begin{array}{c}
 \{p_k+F\}\  ,\ \{h_k-F\}\\
 \{p_k\}\  ,\ \{h_k\} \end{array}\right),
 \end{equation}
where we have used the standard hypergeometric type notation for
ratios of $\Ga$ functions:
 \begin{equation}
\Ga\left(
 \begin{array}{c} a_1\ ,\ \dots\ ,\ a_{\ell}\\
                              b_1\ ,\ \dots\ ,\ b_{j}
 \end{array}\right)
  =  \pl{k=1}{\ell} \Ga(a_k) \cdot \pl{k=1}{j} \Ga(b_k)^{-1}  .
\end{equation}

The exponents $\theta$ in this case are
 \begin{equation}\label{powers-2}
 \begin{array}{l}
 {\dis \theta_{zz}(r)=\left(F^{(z)}_{r,+}\right)^2+\left(F^{(z)}_{r,-}\right)^2,}\num
 {\dis \theta_{+-}(r)=\left(1+F^{(+)}_{r,+}\right)^2+\left(F^{(+)}_{r,-}\right)^2.}
 \end{array}
 \end{equation}
Note that the shift functions $F^{(z/+)}(\lambda)$ enter the
exponents \eqref{powers-2} only in the combination
$F^{(z/+)}_{r}(\lambda)=F^{(z/+)}(\lambda)+r$. It is easy to see
that $F^{(z/+)}_{r}(\lambda)$ are nothing else but the shift
functions of the $(\alpha+r)$-twisted ground states (see
Definition~\ref{Def-aGS}) in the $N$ and $N+1$ sectors respectively.

\begin{rem}
One can write down the results of this section in a more symmetric
form. For this we introduce two constants $f_\mp$ describing the
shift of the utmost parameters $\mu_1$ and $\mu_{N_\kappa}$ of the
$(\alpha+r)$-twisted ground state with respect to the utmost
parameters $\lambda_1$ and $\lambda_{N}$ of the ground state:
 \begin{equation}\label{utmost-shift}
 \begin{array}{l}
 {\dis f_-=\lim_{N,M\to\infty}M\widehat\rho(\lambda_1)(\mu_1-\lambda_1),}\num
 {\dis f_+=\lim_{N,M\to\infty}M\widehat\rho(\lambda_N)(\mu_{N_\kappa}-\lambda_N).}
 \end{array}
 \end{equation}
Then it is easy to see that $f_\pm=F^{(z)}_{r,\pm}$ for the
$\sigma^z$ form factors. However for the $\sigma^+$ form factors
(where $N_\kappa=N+1$) one has $f_-=F^{(+)}_{r,-}$, but
$f_+=1+F^{(+)}_{r,+}$. Then in both cases the exponents
\eqref{powers-2} can be written as $\theta=f_+^2 + f_-^2$.
\end{rem}

It is crucial to compare the results \eqref{powers-2} with the
critical exponents appearing in the asymptotic behavior of the
two-point correlation functions predicted in
\cite{LutP75,Hal80,Hal81a,Hal81b,Aff85,BloCN86,Car84,Car86,Luk99,LukT03,KitKMST09b}.

In the case of $\theta_{zz}(r)$, it follows from \eqref{Z-Th},
\eqref{F-solution}  that $F^{(z)}_r(\lambda)=(\alpha+r)Z(\lambda)$.
Therefore, for $\alpha=0$, we obtain that
$\theta_{zz}(r)=2r^2Z^2(q)$, $|r|=1,2,\dots$. These numbers coincide
with the critical exponents appearing in  the asymptotic behavior of the
$\langle\sigma^z_1\sigma^z_{m+1}\rangle$ correlation function and are associated to the oscillating term with momentum $2rk_F$.

The exponent $\theta_{+-}(r)$ has similar properties. In this case,
$F^{(+)}_r(\lambda)=(\alpha+r-1/2)Z(\lambda)+\phi(\lambda,q)$. It
follows from equations \eqref{Z-Th} and \eqref{prop-phase} that
 \begin{equation}\label{prop-phase1}
 \begin{array}{l}
  {\dis 1+\phi(q,q)=\frac{Z(q)+Z^{-1}(q)}2,}\num
  {\dis \phi(-q,q)=\frac{Z(q)-Z^{-1}(q)}2.}
 \end{array}
\end{equation}
From this we find
 \begin{equation}\label{F+--FS}
 \begin{array}{l}
  {\dis F_{r,+}+1=(\alpha+r)Z(q)+\frac{Z^{-1}(q)}2,}\num
  {\dis F_{r,-}=(\alpha+r)Z(q)-\frac{Z^{-1}(q)}2.}
 \end{array}
\end{equation}
Thus, at $\alpha=0$ we have $\theta_{+-}(r)=Z^{-2}(q)/2+2r^2Z^2(q)$,
$|r|=0,1,\dots$. Again, these numbers coincide with the critical exponents appearing in the asymptotic behavior
of the $\langle\sigma^+_1\sigma^-_{m+1}\rangle$ correlation function and associated to the oscillating term with momentum $2rk_F$.
Thus, we see that the behavior of critical form factors with respect
to the size of the system coincides with the asymptotic behavior of two-point
correlation functions with respect to the lattice distance.

%%%%%%%%%%%%%%%%%%%%%%%%%%%%%%%%%%%%%%%%%%%%%%%%%%%%%%%%%%%%%%%%%%%%%%
%%%%%%%%%%%%%%%%%%%%%%%%%%%%%%%%%%%%%%%%%%%%%%%%%%%%%%%%%%%%%%%%%%%%%%
%%%%%%%%%%%%%%%%%%%%%%%%%%%%%%%%%%%%%%%%%%%%%%%%%%%%%%%%%%%%%%%%%%%%%%
%%%%%%%%%%%%%%%%%%%%%%%%%%%%%%%%%%%%%%%%%%%%%%%%%%%%%%%%%%%%%%%%%%%%%%
%%%%%%%%%%%%%%%%%%%%%%%%%%%%%%%%%%%%%%%%%%%%%%%%%%%%%%%%%%%%%%%%%%%%%%
%%%%%%%%%%%%%%%%%%%%%%%%%%%%%%%%%%%%%%%%%%%%%%%%%%%%%%%%%%%%%%%%%%%%%%

\section{Form factors and scalar products\label{FFaSP}}

The solution of the quantum inverse scattering problem enables us to
express the local spin operators in terms of the entries of the
monodromy matrix \cite{KitMT99,MaiT00}:
 \begin{equation}\label{FCtab}
 \sigma^s_m
  ={\cal T}^{m-1} (-i\zeta/2) \cdot
   \tr\big(T (-i\zeta/2)\, \sigma^s\big)
    \cdot {\cal T}^{-m} (-i\zeta/2) .
 \end{equation}
In the \textit{l.h.s.} of this expression, the symbol $\sigma^s_m$ ($s=\pm, z$) denotes the corresponding local spin operator
at site $m$, whereas the symbol $\sigma^s$ appearing in the \textit{r.h.s.} should be
understood as a $2\times 2$ Pauli matrix multiplying the $2\times 2$
monodromy matrix \eqref{Mon-Mat}.

Using \eqref{FCtab}, one can reduce the computation of the form
factors of local spin operators in the finite XXZ chain to the one
of the scalar products \cite{KitMT99}. We have
 \begin{multline}\label{m-el-sz}
 {\cal F}^{(z)}_{\psi' \psi_g}(m)
 =-e^{i(m-1)\sum\limits_{j=1}^N [p_0(\mu_{\ell_j})-p_0(\lambda_j) ]}
 \bigg(e^{i\sum\limits_{j=1}^N [p_0(\mu_{\ell_j})-p_0(\lambda_j) ]}-1\bigg) \\
 \times
 \Bigl.\frac{\partial}{\partial\alpha}\frac{\langle\psi_\kappa(\{\mu\})|
 \psi(\{\lambda\})\rangle}{\pi i\|\psi_\kappa(\{\mu\})\|\cdot \|\psi(\{\lambda\})\|}\Bigr|_{\alpha=0},
 \end{multline}
which is nonzero only if $N_{\kappa}=N$.  The form-factor ${\cal
F}^{(z)}_{\psi_g \psi'}$ is obtained by complex conjugation ${\cal
F}^{(z)}_{\psi_g \psi'}(m)= \pa{{\cal F}^{(z)}_{\psi'\psi_g}(m)
}^{*}$. Similarly, one obtains
 \begin{align}
 {\cal F}^{(-)}_{\psi'\psi_g}(m)&= (-1)^{N+m+1}\,
 e^{i(m-1)\!\sum\limits_{j=1}^{N_\kappa} p_0(\mu_{\ell_j})-im\!\sum\limits_{j=1}^{N}p_0(\lambda_j) }
 \frac{\langle\psi_\kappa(\{\mu\})|B\big({\textstyle-\frac{i\zeta}2}\big)|
 \psi(\{\lambda\})\rangle}{ a(-i\frac{\zeta}{2})\, \|\psi_\kappa(\{\mu\})\|\cdot \|\psi(\{\lambda\})\|}\Bigr|_{\alpha=0},
 \label{m-el-s-}\\
 {\cal F}^{(+)}_{\psi_g\psi'}(m)&= (-1)^{N+m}\,
 e^{i(m-1)\!\sum\limits_{j=1}^{N} p_0(\lambda_j)-im\!\sum\limits_{j=1}^{N_\kappa}p_0(\mu_{\ell_j}) }
 \frac{\langle\psi(\{\lambda\})|C\big({\textstyle-\frac{i\zeta}2}\big)|
 \psi_\kappa(\{\mu\})\rangle}{ a(-i\frac{\zeta}{2})\,\|\psi_\kappa(\{\mu\})\|\cdot \|\psi(\{\lambda\})\|}\Bigr|_{\alpha=0},
 \label{m-el-s+}
 \end{align}
which are nonzero only if $N_{\kappa}=N+1$. In \eqref{m-el-s-} and
\eqref{m-el-s+},  $a(\nu)=\sinh^{M}\pa{\nu-i\tf{\zeta}{2}}$ denotes
the eigenvalue of the operator $A(\nu)$ on the reference state
which, in the case of the spin-1/2 chain, is the ferromagnetic state
with all spins up. In the following, we will also use the notation
$d(\nu)=\sinh^{M}\pa{\nu+i\tf{\zeta}{2}}$ for the eigenvalue of
$D(\nu)$ on this reference state.

\begin{rem}
Although it is possible to set $\alpha=0$ directly in
\eqref{m-el-s-} and \eqref{m-el-s+}, we prefer to take the limit
$\alpha=0$ only in the very end of the calculations  to study the
properties of the scalar products for general $\alpha$.
\end{rem}

In all the above scalar products, one of the states is an eigenstate
of the twisted transfer matrix. There exists an explicit determinant
representations for such scalar products  and the associated norms:
\begin{prop}
\label{proposition produits scalaire Fredholm}
Let $\paa{\la}_1^{N}$ satisfy the ground state Bethe  equations
\eqref{BE} and $\paa{\mu}_1^{N_{\kappa}}$ solve the system of
$\alpha$-twisted Bethe Ansatz equations \eqref{TBE-lj}. Then the
following representations for the scalar products hold:
 \begin{multline}\label{2-l-rep}
 \langle \psi_\kappa(\{\mu\})|\psi(\{\lambda\})\rangle
 = \delta_{N_\kappa,N} \cdot
 \prod_{a,b=1}^N
 \frac{\sinh(\mu_{\ell_a}-\lambda_{b}-i\zeta)}{\sinh(\lambda_a-\mu_{\ell_b})}\cdot
 \prod_{j=1}^N\left\{d(\mu_{\ell_j})\, d(\lambda_j)
\pac{e^{2i\pi \wh{F}\pa{\la_j}}-1} \right\}
  \num
 \times
  \frac{1-\kappa}{1-e^{2\pi i\widehat F(-q)}}\cdot \prod_{a=1}^{N}
 \frac {\sinh(q+\lambda_a-i\zeta)}{\sinh(q-\mu_{\ell_a}-
 i\zeta)}\cdot
  \det_{\Ga_q}\left[ I+\frac{1}{2\pi i} \widehat U^{\pa{z}}(w,w') \right] ,
  \end{multline}
\begin{multline}
\label{Ps-B}
 \langle \psi_\kappa(\{\mu\})|\, B(-i \zeta/2)\, |\psi(\{\lambda\})\rangle=
\delta_{N_\kappa,N+1}\cdot a(-i \zeta/2) \sinh\pa{-i\zeta}
\pl{j=1}{N}\paa{ a(\la_j)  \pac{1-e^{2i\pi \wh{F}\pa{\la_j}}} }
          \\
\times         \f{ \pl{b=1}{N} \sinh(\la_b-i\tf{\zeta}{2})  }{ \pl{b=1}{N+1} \sinh(\mu_{\ell_b}+i\tf{\zeta}{2})  } \,
      \pl{a=1}{N+1}\left\{ d(\mu_{\ell_a}) \pl{b=1}{N}\frac{  \sinh(\mu_{\ell_a}-\la_b-i\zeta) }
 {\sinh(\mu_{\ell_a}-\la_{b})  }\right\} \cdot \det_{\Ga_q}\!\pac{ I+
\frac{1}{2\pi i} \wh{U}^{\pa{+}}\pa{w,w'} } .
\end{multline}
In these expressions $\widehat F(\lambda)$ \eqref{SP-shiftF-mod}
depends on whether we consider the $N$ sector or the $(N+1)$ sector
(see \eqref{TBE-cf}). The integral operators $I+\frac1{2\pi i}
\widehat U^{(z)}$ and $I+\frac1{2\pi i}\widehat U^{(+)}$ act on a
closed anti-clockwise oriented contour $\Ga_q$ surrounding the
interval $[-q,q]$ where the ground state roots
$\paa{\la_a}_{a=1}^{N}$ condensate and containing no other
singularity of the kernels. The last ones have the form
 \begin{equation}\label{2-U1}
 \widehat U^{\pa{z}}(w,w')=
 -\prod\limits_{a=1}^{N}\frac{\sinh(w-\mu_{\ell_a})\sinh(w-\lambda_a+i\zeta)}
 {\sinh(w-\lambda_a)\sinh(w-\mu_{\ell_a}+i\zeta)}
 \cdot
 \frac{K_\kappa(w-w')-K_\kappa(-q-w')}{1-e^{2\pi i\widehat
 F(w)}},
 \end{equation}
 \begin{equation}\label{2-U2}
 \wh{U}^{(+)}(w,w')=\prod\limits_{a=1}^{N+1}\frac{\sinh(w-\mu_{\ell_a})}
 {\sinh(w-\mu_{\ell_a}+i\zeta)}
  \prod\limits_{a=1}^{N}\frac{\sinh(w-\lambda_a+i\zeta)}
 {\sinh(w-\lambda_a)}\cdot
 \frac{{\cal K}_\kappa(w,w')}{1-e^{2\pi i\widehat F(w)}},
 \end{equation}
where $K_\kappa$ and $\mathcal{K}_\kappa$ are given respectively by
\eqref{2-Kk} and \eqref{K-2}.
\end{prop}
The representation \eqref{2-l-rep} was obtained in
\cite{Sla89,Sla95,KitMT99,KitKMST09b}. The representation
\eqref{Ps-B} is derived in appendix~\ref{S-det-rep}.

We also recall the finite size determinant representations for the norms of Bethe states:
\begin{prop}\cite{GauMT81,Kor82,Gau83}
\label{proposition norme}
Let $\mu_{\ell_1},\dots,\mu_{\ell_{N_{\kappa}}}$ satisfy the system \eqref{TBE-cf}. Then
 \begin{multline}\label{norm-Theta}
 \langle \psi_\kappa(\{\mu\})|\psi_\kappa(\{\mu\})\rangle
 = (-1)^{N_{\kappa}}
 \prod_{j=1}^{N_{\kappa}}  \bigl[2\pi iM \, \wh\rho_\kappa(\mu_{\ell_j}) \, a(\mu_{\ell_j}) \, d(\mu_{\ell_j})\bigr] \;
 \frac{\prod\limits_{a,b=1}^{N_\kappa}\sinh(\mu_{\ell_a}-\mu_{\ell_b}-i\zeta)}
        {\prod\limits_{a,b=1\atop{a\ne b}}^{N_{\kappa}}\sinh(\mu_{\ell_a}-\mu_{\ell_b})}\\
 \times
 \det_{N_{\kappa}} \Theta_{jk}^{(\mu)},
 \end{multline}
where
 \begin{equation}\label{2-Gjk}
 \Theta_{jk}^{(\mu)}=\delta_{jk}+\frac{K(\mu_{\ell_j}-\mu_{\ell_k})}{2\pi M\wh\rho_\kappa(\mu_{\ell_k})}.%,
 %\qquad
 %2\pi M \wh\rho_\kappa(\mu)=-i\log'\frac{a(\mu)}{d(\mu)}-\sum_{a=1}^{N_{\kappa}} K(\mu-\mu_{\ell_a}).
%
 \end{equation}
\end{prop}

One can see from another representation for the scalar product \eqref{scal-prod} that, when  $\Im\alpha=0$,
 \begin{equation}\label{conj-1}
 \langle\psi_\kappa(\{\mu\})|\psi(\{\lambda\})\rangle=
 \Bigl(\langle\psi_\kappa(\{\mu\})|\psi(\{\lambda\})\rangle\Bigr)^*,
 \end{equation}
and
 \begin{equation}\label{sc-prod-CC}
 \langle\psi(\{\la\})| C ({\textstyle-\frac{i\zeta}2}) | \psi_\kappa(\{\mu\})\rangle=
 \kappa^{-1}e^{i\sum_{j=1}^{N_\kappa}p_0(\mu_{\ell_j})+i\sum_{j=1}^{N}p_0(\lambda_j)}
 \Bigl(\langle\psi_\kappa(\{\mu\})| B({\textstyle-\frac{i\zeta}2})|
 \psi(\{\la\})\rangle\Bigr)^*\;.
 \end{equation}

For the calculation of  the two-point correlation functions
$\langle\sigma^z_m\sigma^z_{m'}\rangle$ (resp. $\langle\sigma^+_m\sigma^-_{m'}\rangle$),
we need actually to sum up
the products ${\cal F}^{(z)}_{\psi_g\psi'}(m)\,{\cal F}^{(z)}_{\psi'\psi_g}(m')$
(resp.  ${\cal F}^{(+)}_{\psi_g\psi'}(m)\,{\cal F}^{(-)}_{\psi'\psi_g}(m')$)
over all eigenstates $\ket{\psi'}$.
We have the following result:
\begin{prop}
The products of two form factors  can be written in terms of the former scalar products as
 \begin{equation}\label{m-el-sigma}
 {\cal F}^{(z)}_{\psi_g\psi'}(m')\cdot{\cal F}^{(z)}_{\psi'\psi_g}(m)
 =e^{i(m-m'){\wh{\cal P}}_{ex}} \, \frac{2\sin^2\left(\frac{\wh{\cal P}_{ex}}2\right)}{\pi^2} \cdot
  \left.\frac{\partial^2}{\partial\alpha^2}
 S_N^{z}
 \right|_{\alpha=0} \, ,
 \end{equation}
 \begin{equation}\label{f-f-SP-1}
 {\cal F}^{(+)}_{\psi_g\psi'}(m') \cdot {\cal F}^{(-)}_{\psi'\psi_g}(m)=\left.(-1)^{m-m'}e^{i(m-m')\wh{\cal
 P}_{ex}} \cdot S_N^{+}\right|_{\alpha=0} \, ,
 \end{equation}
where
 \begin{equation}\label{S}
 S_N^{z}
 =\left|\frac{ \langle\psi_\kappa(\{\mu\})|
 \psi(\{\lambda\})\rangle   }
 {\|\psi_{\kappa}(\{\mu\})\|\cdot\|\psi(\{\lambda\})\|}\right|^2,
 \end{equation}
 \begin{equation}\label{SP-ff}
 S_N^{+}= -\frac{e^{-2\pi i\alpha}}{ a^2(-i\zeta/2)}
 \left|\frac{ \langle\psi_\kappa(\{\mu\})|B\big({\textstyle-\frac{i\zeta}2}\big)|
 \psi(\{\lambda\})\rangle   }
 {\|\psi_{\kappa}(\{\mu\})\|\cdot\|\psi(\{\lambda\})\|}\right|^2.
 \end{equation}
In \eqref{m-el-sigma} and \eqref{f-f-SP-1}, $\widehat{\cal P}_{ex}$
denotes the  relative excitation momentum of the excited states in
the $N_{\kappa}$ sector in respect to the ground state in the $N$
sector,
 \begin{equation}\label{Mom-ex}
 \wh{\cal P}_{ex}=\sum_{j=1}^{N_\kappa}
 p_0(\mu_{\ell_j})-\sum_{j=1}^{N}p_0(\lambda_j).
 \end{equation}
\end{prop}

\begin{rem}
Here, we did not
consider the product ${\cal F}^{(-)}_{\psi_g\psi'}(m)\, {\cal
F}^{(+)}_{\psi'\psi_g}(m')$, since the equal-time correlation function
$\langle\sigma^-_m\sigma^+_{m'}\rangle$ can be obtained from
$\langle\sigma^+_m\sigma^-_{m'}\rangle$ by the replacement $m\to m'$.
\end{rem}

The calculation of the thermodynamic limit of $\widehat{\cal
 P}_{ex}$ is a relatively simple problem. Indeed, it follows from
\eqref{BE-cf}, \eqref{TBE-cf} that
 \begin{equation}\label{blim-Mom-ex}
 \frac{\wh{\cal P}_{ex}}{2\pi} =\alpha\frac{N_\kappa}M
 +\sum_{j=1}^{N_\kappa}\hat\xi_\kappa(\mu_{\ell_j})-
 \sum_{j=1}^{N}\hat\xi(\lambda_{j})
 +\frac{N^2+N-N_\kappa^2-N_\kappa}{2M}=\alpha\frac{N_\kappa}M
 +\sum_{k=1}^{n}\bigl(\hat\xi_\kappa(\mu_{p_k})-\hat\xi_\kappa(\mu_{h_k})\bigr).
 \end{equation}
Here we have used the antisymmetry of the bare phase
$\vartheta(-\lambda)=-\vartheta(\lambda)$ and
$\hat\xi_\kappa(\mu_{\ell_j})=\ell_j/M$, $\hat\xi(\lambda_{j})=j/M$.
Then, using \eqref{0-Dmom}, we obtain that the thermodynamic limit ${\cal P}_{ex}$ of $\wh{\cal P}_{ex}$ is given by the expression \eqref{mom-ex}.

The non-trivial problem is the computation of the thermodynamic limit of the scalar products $S_N^{z}$
and $S_N^{+}$. This problem will be treated in the remaining part of the article.

%%%%%%%%%%%%%%%%%%%%%%%%%%%%%%%%%%%%%%%%%%%%%%%%%%%%%%%%%%%%%%%%%%%%%%
%%%%%%%%%%%%%%%%%%%%%%%%%%%%%%%%%%%%%%%%%%%%%%%%%%%%%%%%%%%%%%%%%%%%%%
%%%%%%%%%%%%%%%%%%%%%%%%%%%%%%%%%%%%%%%%%%%%%%%%%%%%%%%%%%%%%%%%%%%%%%
%%%%%%%%%%%%%%%%%%%%%%%%%%%%%%%%%%%%%%%%%%%%%%%%%%%%%%%%%%%%%%%%%%%%%%
%%%%%%%%%%%%%%%%%%%%%%%%%%%%%%%%%%%%%%%%%%%%%%%%%%%%%%%%%%%%%%%%%%%%%%
%%%%%%%%%%%%%%%%%%%%%%%%%%%%%%%%%%%%%%%%%%%%%%%%%%%%%%%%%%%%%%%%%%%%%%

\section{Thermodynamic limit of $S_N^{z}$\label{ThDL-Sz}}

In the paper  \cite{KitKMST09c} we  studied  the thermodynamic limit
of the scalar product $S_N^{z}$ in the particular case where the
state $|\psi_\kappa(\{\mu\})\rangle$ was the $\alpha$-twisted ground
state in the $N$ sector. In the general case, when the excited state
contains $n$ particles and $n$ holes, the result also depends on
their rapidities $\{\mu_p\}$ and $\{\mu_h\}$.
As we have already announced, such dependence  is not sufficient to
characterize completely the thermodynamic limit of the form factor.
Indeed, the latter decomposes into a product of a smooth and a
discrete part. The smooth part can actually be completely described
in terms of the rapidities of the particles and holes. However, for
the description of the discrete part one should use also the integer
numbers $\{p_a\}$ and $\{h_a\}$.

In the first part of this section, we explicitly factorize $S_N^{z}$
into the aforementioned product. Then, we investigate the
thermodynamic limit of the smooth part, postponing the more
complicated analysis of the discrete part until
subsection~\ref{TDlimDP}.

Note that in this section we deal only with the  excited states in
the $N$ sector. Therefore the thermodynamic limit of the  shift
function $\widehat F(\lambda)$ is equal to $F^{(z)}(\lambda)$ given
by \eqref{F-solution}. However, in order to lighten notations we
omit the superscript $(z)$ throughout this section, denoting the
limiting value of the shift function simply by $F(\lambda)$.

\subsection{Representation of the scalar product
$S_N^{z}$\label{WWS}}

Assume that the excited state $|\psi_\kappa(\{\mu\})\rangle$ is an
excited state in the $N$ sector containing $n$ particles with
rapidities $\{\mu_{p_a}\}_{a=1}^{n}$ and $n$ holes with rapidities
$\{\mu_{h_a}\}_{a=1}^{n}$. One can fairly expect that the limiting
value of $S_N^{z}$  depends on the thermodynamic limit of these
rapidities $\{\mu_p\}$ and $\{\mu_h\}$, \textit{i.e.} $\lim S_N^{z}=
S_n^{z}(\{\mu_p\},\{\mu_h\})$.

Using Propositions \ref{proposition produits scalaire Fredholm} and
\ref{proposition norme} as well as \eqref{S}, it is readily seen
that $S_N^{z}$ has the following representation:
 \begin{equation}\label{S-prom}
 S_N^{z}(\{\mu_p\},\{\mu_h\})=
 {\cal A}_N^{(z)}(\{\mu_p\},\{\mu_h\})
 \cdot D_N^{(z)}(\{\mu_p\},\{\mu_h\})\cdot
\exp\paa{C_{N}^{(z)}(\{\mu_p\},\{\mu_h\})} ,
 \end{equation}
where
 \begin{equation}\label{W}
 C_{N}^{(z)}(\{\mu_p\},\{\mu_h\})
 =\sum_{a,b=1}^N\log
   \frac{\sinh(\lambda_a-\mu_{\ell_b}-i\zeta) \, \sinh(\mu_{\ell_b}-\lambda_a-i\zeta)}
           {\sinh(\lambda_a-\lambda_b-i\zeta) \,
           \sinh(\mu_{\ell_a}-\mu_{\ell_b}-i\zeta)},
 \end{equation}
 \begin{equation}\label{6-DNk}
 D_N^{(z)}(\{\mu_p\},\{\mu_h\})=\left(\det_N\frac1{\sinh(\mu_{\ell_j}-\lambda_k)}\right)^2\cdot
 \prod\limits_{j=1}^N\frac{\sin^2\pi\widehat
 F(\lambda_j)} {\pi^2M^2 \, \widehat\rho(\lambda_j) \,
 \widehat\rho_\kappa(\mu_{\ell_j})},
 \end{equation}
and
 \begin{equation}\label{6-ev-phiC}
 {\cal A}_N^{(z)}(\{\mu_p\},\{\mu_h\})
    =\left|\frac{\sin\pi\alpha }{\sin\pi \widehat F(-q)}  \prod_{a=1}^{N}
 \frac {\sinh(q+\lambda_a-i\zeta)}{\sinh(q-\mu_{\ell_a}-
 i\zeta)}\right|^2   \frac{\left|\det_{\Ga_q}\Big[I+\frac1{2\pi i}\widehat U^{\pa{z}}(w,w')\Big]\right|^2}
 {  \det_N\Theta_{jk}^{(\lambda)}
 \cdot   \det_N\Theta_{jk}^{(\mu)}}.
 \end{equation}
We remind here that $\wh{U}^{\pa{z}}$ has  been defined in
\eqref{2-U1} and that the $N\times N$ matrices
$\Theta_{jk}^{(\lambda)}$ and $\Theta_{jk}^{(\mu)}$ are
 \begin{equation}\label{2-Gjk-bis}
 \Theta_{jk}^{(\lambda)}=\delta_{jk}+\frac{K(\lambda_j-\lambda_k)}{2\pi M\widehat\rho(\lambda_k)},
 \qquad
 \Theta_{jk}^{(\mu)}=\delta_{jk}+\frac{K(\mu_{\ell_j}-\mu_{\ell_k})}{2\pi
 M\widehat\rho_\kappa(\mu_{\ell_k})} \;.
 \end{equation}

\subsection{Thermodynamic limit of the smooth part\label{TL-SP-z}}

The smooth part of the scalar product consists of factors ${\cal
A}_N^{(z)}$ and $C_{N}^{(z)}$. Their  limits  can be computed
exactly in the same manner as in \cite{KitKMST09c}. We set
 \begin{equation}\label{def-CnAn}
 \begin{array}{l}
 C_{n}^{(z)}(\{\mu_p\},\{\mu_h\})=\lim\limits_{N,M\to\infty}C_{N}^{(z)}(\{\mu_p\},\{\mu_h\}),\\
 {\cal A}_n^{(z)}(\{\mu_p\},\{\mu_h\})= \lim\limits_{N,M\to\infty}{\cal
 A}_N^{(z)}(\{\mu_p\},\{\mu_h\}).
 \end{array}
 \end{equation}

We illustrate the main idea behind the calculation of $C_{n}^{(z)}$
and ${\cal A}_n^{(z)}$ on the following toy-example. Assume that
$f(\lambda)\in C^1(\R)$, and consider the computation of the
thermodynamic limit of the sum
 \begin{equation}\label{Sf}
 S_f=\sum_{j=1}^N\bigl[f(\mu_{\ell_j})-f(\lambda_j)\bigr] \, .
 \end{equation}
To compute this limit one can decompose the sum as follows
 \begin{equation}\label{Lim-Sf}
 S_f={\sum_{j=1}^N}\bigl[f(\mu_{j})-f(\lambda_j)\bigr]+
 \sum_{j=1}^n\bigl[f(\mu_{p_j})-f(\mu_{h_j})\bigr]  \to
 \sum_{j=1}^n\bigl[f(\mu_{p_j})-f(\mu_{h_j})\bigr]+
 \int\limits_{-q}^q \! F(\lambda)f'(\lambda)\,d\lambda.
 \end{equation}
There we have used the leading behavior in $M^{-1}$ \eqref{shift-TD}
for the spacing between $\mu_j$ and $\la_j$. It is convenient to
introduce a modified shift function by
 \begin{equation}\label{F-mod}
 F_{mod}(\lambda)=F(\lambda)\chi_{{}_{[-q,q]}}(\lambda)+\sum_{j=1}^n
 \left(\chi_{{}_{]-\infty,\mu_{p_j}]}}(\lambda) - \chi_{{}_{]-\infty,\mu_{h_j}]}}(\lambda)\right),
 \end{equation}
where $\chi_{[a,b]}$ is the characteristic function of the interval
$[a,b]$. Then equation \eqref{Lim-Sf} can be recast simply as
 \begin{equation}\label{Lim-Sf-1}
 S_f\to
  \int\limits_{-\infty}^\infty \! F_{mod}(\lambda) \, f'(\lambda)\,d\lambda.
 \end{equation}
Therefore, provided that one extends the integration contours to the
whole real axis and replaces $F$ with $F_{mod}$, one formally
reduces the computations to the case considered in \cite{KitKMST09c}
where one was dealing with an excited state
$|\psi_\kappa(\{\mu\}\rangle$ having no particles or holes. From
this observation, the results of section~\ref{SFinRes} follow
straightforwardly (by a mere replacement of the shift function by
$F_{mod}$) from those of \cite{KitKMST09c}. In this way, starting
from the equations \eqref{W}, \eqref{6-ev-phiC} we arrive at
\eqref{Cnz}, \eqref{Anz}.

Thus, the limits  ${\cal A}_n^{(z)}$ and $C_{n}^{(z)}$ are well
defined for arbitrary positions of particles and holes. They depend
 on $\{\mu_{p}\}$ and $\{\mu_{h}\}$ only. In particular, they do not depend on the
underlying integer numbers.

At this stage it is interesting  to consider the limiting case
where, in the thermodynamic limit, all rapidities of particles and
holes condensate on the Fermi boundaries.

\begin{cor}
Let $\ket{\psi_\kappa(\{\mu\})}$ belongs to the $\mathbf{P}_{r}$
class as in \eqref{Nak-mom}, and denote
 \begin{equation}\label{not-r}
 \begin{array}{l}
 C_{n,r}^{(z)}=C_{n}^{(z)}(\{+q\}_{n_p^+}\cup\{-q\}_{n_p^-},
 \{+q\}_{n_h^+}\cup\{-q\}_{n_h^-}),\\
 {\cal A}_{n,r}^{(z)}={\cal A}_{n}^{(z)}(\{+q\}_{n_p^+}\cup\{-q\}_{n_p^-},
 \{+q\}_{n_h^+}\cup\{-q\}_{n_h^-}) \; ,
 \end{array}
 \end{equation}
where subscripts show the number of elements in the corresponding
subsets. Let $F_r(\la)=F(\la)+r$ and $\mathbf{F}_r(w)$ be the
$i\pi$-periodic Cauchy transform of $F_r(\la)$. Then the coefficient
$C_{n,r}^{(z)}$ takes the form
\begin{equation}
C_{n,r}^{(z)}= C_{0}[F_r ],
  \end{equation}
where the functional $C_{0}[F]$ is defined in \eqref{C0}. The
coefficient ${\cal A}_{n,r}^{(z)}$ becomes
 \begin{equation}\label{Anz-r}
  {\cal A}_{n,r}^{(z)}
    =\left(\frac{\sin\pi\alpha }{\sin\pi F(-q)}\right)^2
 \left|e^{-2\pi i\mathbf{F}_r(-q+i\zeta)}\right|^2
   \left|\frac{\det\big[I+\frac1{2\pi i}U_r^{(z)}(w,w')\big] }
    {\det \big[ I+\frac1{2\pi}K \big] }\right|^2,
  \end{equation}
where
 \begin{equation}\label{U-qz-r}
 U_r^{(z)}(w,w')=-e^{2\pi i\bigl(\mathbf{F}_r(w)-\mathbf{F}_r(w+i\zeta)\bigr)} \frac{K_\kappa(w-w')-K_\kappa(-q-w')}
 {1-e^{2\pi iF(w)}} \;.
 \end{equation}
\end{cor}

\proof This follows straightforwardly from
$F_{mod}(\lambda)=F_r(\lambda)\, \chi_{[-q,q]}(\lambda)$.
\qed

\vspace{3mm} Using \eqref{Z-Th}, \eqref{F-solution} we find that
 \begin{equation}\label{Fr-Z}
 F_r(\lambda) =\left(r+\alpha\right)Z(\lambda),
 \end{equation}
and we see that up to a replacement $\alpha+r\to\alpha$ our results
coincide with the ones obtained in \cite{KitKMST09c}. This
coincidence is not accidental.

In the thermodynamic limit, the class $\mathbf{P}_{r}$ contains an infinite number of states.
Indeed, knowing that a particle or hole's rapidity equals to $\pm q$ in this limit
does not allow to fix the corresponding integers $p$ (resp. $h$)
unambiguously.  For instance, knowing that $\mu_p=q$ only allows one
to say that $\tf{p}{M}\to D$. The latter is satisfied as long as one
chooses $p=N+u_M$ with $u_M=\e{o}\pa{M}$, but arbitrary otherwise.

It is clear that quantities having well defined thermodynamic limits
(such as $C_{n,r}^{(z)}$ and ${\cal A}_{n,r}^{(z)}$), must only
depend on the macroscopic realization $\{\mu_p\}$ and $\{\mu_h\}$ of
the excited state and not on the microscopic quantum numbers $\{p\}$
and $\{h\}$ leading to such a macroscopic configuration of the
rapidities. Therefore, the calculation of their thermodynamic limit
in the case of an excited state belonging to the $\mathbf{P}_{r}$
class can be done by choosing any of its representative. One of
these representatives is given by the $(\alpha+r)$-twisted ground
state. Indeed, set $n_p^-=n_h^+=0$ and choose the integers
describing the holes and particles according to
 \begin{equation}\label{integers}
 h_k=k, \qquad p_k=N+k, \qquad k=1,\dots,r \; .
 \end{equation}
The excited state is thus described by the set of integers,
 \begin{equation}\label{integers1}
 \ell_j=j\quad\mbox{for}\quad j=r+1,\dots,N, \quad \e{and} \quad  \ell_j=N+j\quad\mbox{for}\quad j=1,\dots,r\; .
 \end{equation}
Observe now that if we replace $\alpha$ by $\alpha+r$ in the
\textit{r.h.s.} of \eqref{TBE-lj}, then  up to a re-ordering of the
equations, we obtain the same set of integers \eqref{integers1}.
Thus, the above representative of the class $\mathbf{P}_{r}$
coincides with the $(\alpha+r)$-twisted ground state. Therefore the
smooth part of all the critical form factors belonging to the
$\mathbf{P}_{r}$ class coincides with the one of the form factor of
the $\sigma^z$ operator taken between the ground state and the
$(\alpha+r)$-twisted ground state.

\subsection{Thermodynamic limit of the discrete part\label{TDlimDP}}

As we have seen, the modified shift function \eqref{F-mod} can be
used for the calculation of the thermodynamic limits of sums
(products, determinants) in the case when we deal with smooth
functions of the rapidities of the excited state. One cannot apply
this method in the case of the factor
$D_N^{(z)}(\{\mu_p\},\{\lambda_h\})$ as it depends on the Cauchy
determinant of $\{\lambda\}$ and $\{\mu\}$. In the thermodynamic
limit, certain entries of this matrix become divergent and more care
is needed for the calculation of this limit.
\begin{prop}
\label{Proposition asymptotique Dz}

The discrete part $D_N^{(z)}$ behaves in the thermodynamic as
 \begin{multline}\label{DNk-new}
 D_N^{(z)}(\{\mu_p\},\{\mu_h\})=  M^{-F_+^2-F_-^2} \cdot  E_0\pa{\paa{\mu_h,\mu_p};\paa{h_a,p_a}}
 \cdot \prod_{k=1}^{n} H_N\pa{\mu_{h_k},h_k} \\
\times  \prod_{k=1}^{n} P_N\pa{\mu_{p_k},p_k}  \cdot
 \mc{D}^{\pa{z}}[F] \cdot  \pa{1+\e{O}\paf{\log M}{M}}  \; .
 \end{multline}
The factor $E_0$ depends explicitly on the integers $\{h_a\}$ and $\{ p_a \}$ which are the quantum numbers  defining the excited state
 \begin{equation}\label{E0}
 E_0=\frac{\prod\limits_{j,k=1\atop{j\ne
 k}}^n(h_j-h_k) (p_j-p_k) \,
 \varphi(\mu_{h_j},\mu_{h_k})\,
 \varphi(\mu_{p_j},\mu_{p_k})}
 {\prod\limits_{j,k=1}^n (p_j-h_k)^2 \, \varphi^2(\mu_{p_j},\mu_{h_k})},
 \end{equation}
where $\vp\pa{\la,\mu}=  \tf{\pa{2\pi \sinh(\la-\mu)}}{\pa{p(\la)-p(\mu)}}$ and
$p(\lambda)$ is the dressed momentum \eqref{0-Dmom}.
The factors $P_N$ and $H_N$ also depend on the quantum numbers defining the holes and the particles
 \begin{equation}\label{Ekp}
P_N\pa{\mu_{p_k},p_k}= \f{ e^{J[F^{\pa{z}}]\pa{\mu_{p_k}}}
}{\rho\pa{\mu_{p_k}}} \;
 \Gamma^2\! \left(
 \begin{array}{c}
 p_k\  ,\ p_k-N+ F(\mu_{p_k})\\
 p_k+ F(\mu_{p_k})\  ,\ p_k-N
 \end{array}\right)  ,
 \end{equation}
 \begin{equation}\label{Ekh}
 H_N\pa{\mu_{h_k},h_k}= \f{ \sin^2(\pi F(\mu_{h_k})) }{e^{J[F^{\pa{z}}]\pa{\mu_{h_k}}} \pi^2 \rho\pa{\mu_{h_k}}} \;
 \Gamma^2 \! \left(
 \begin{array}{c}
 h_k+ F(\mu_{h_k}) \  ,\ N+1-h_k- F(\mu_{h_k})\\
 h_k\  ,\ N+1-h_k
 \end{array}\right)  .
 \end{equation}
There we have introduced the functional $J$ which is defined in
terms of the dressed momentum $p(\lambda)$ and acts on the shift
function $F$:
 \begin{equation}
\label{fonction J} J\pac{F}\pa{\om}= 2 \Int{-q}{q} \paa{ F\pa{\la}
 \partial_\lambda \log \vp(\la,\om) + \f{
F\pa{\la}-F\pa{\om}}{p\pa{\la}-p\pa{\om}} p^{\prime}\pa{\la} } d \la
\, .
 \end{equation}
Finally, we remind that the functional $\mc{D}^{\pa{z}}$ is given in \eqref{fonctionnelle Dz}
\end{prop}

\begin{rem}
The definition of $P_N\pa{\mu_{p_k},p_k}$ is given in the  case
where the rapidities of all particles  are to the right of the Fermi
zone ( \textit{ie} $p_k>N$). If some
 particles have their rapidities to the left of the Fermi zone ($p_k\le 0$), then
the corresponding arguments of the $\Gamma$-functions in \eqref{Ekp}
become negative integers. The aforementioned formula remains however
valid provided that the arguments of these $\Gamma$-function are
understood as limits
 \begin{equation}\label{lim-G}
 \frac{\Gamma(p_k)}{\Gamma(p_k-N)}=\lim_{\varepsilon\to 0}
  \frac{\Gamma(p_k+\varepsilon)}{\Gamma(p_k-N+\varepsilon)}=
  (-1)^N\frac{\Gamma(N+1-p_k)}{\Gamma(1-p_k)}\; .
  \end{equation}
\end{rem}

\medskip

\proof Following the strategy applied in \cite{KitKMST09c}, we
multiply and divide the original Cauchy determinant by the Cauchy
determinant of the counting functions $\widehat
\xi_\kappa(\lambda)$. Let
 \begin{equation}\label{XXZ-Uniform}
 \wh\varphi(\lambda,\mu)=\frac{\sinh(\lambda-\mu)}{\wh\xi(\lambda)-\wh\xi(\mu)},\qquad
  \wh\varphi_\kappa(\lambda,\mu)=\frac{\sinh(\lambda-\mu)}{\wh\xi_\kappa(\lambda)-\wh\xi_\kappa(\mu)}.
 \end{equation}
After some algebra, we recast
$D_N^{(z)}(\{\mu_p\},\{\lambda_h\})$ into the following product:
 \begin{equation}\label{DNk-new-2}
 D_N^{(z)}(\{\mu_p\},\{\lambda_h\})=\wh{E}_0
\cdot \prod_{k=1}^{n} \wh{H}_k \cdot \prod_{k=1}^{n} \wh{P}_k
\cdot D_{N,0} \, .
 \end{equation}
Here $\wh{E}_0$ depends only on the particle/hole rapidities and the
corresponding integers:
 \begin{equation}\label{E0-bis}
 \wh{E}_0=
 \frac{\prod\limits_{j,k=1\atop{j\ne k}}^n(h_j-h_k)(p_j-p_k)\, \wh\varphi_\kappa(\mu_{h_j},\mu_{h_k}) \,
 \wh\varphi_\kappa(\mu_{p_j},\mu_{p_k})}
 {\prod\limits_{j,k=1}^n(p_j-h_k)^2 \,  \wh{\varphi}_\kappa^{\,2} \,
 (\mu_{p_j},\mu_{h_k})}.
 \end{equation}
The factors $\wh{P}_k$ depend on the rapidities $\mu_{p_k}$,
integers $p_k$, the ground state parameters $\{\lambda\}$, and the
roots $\mu_j$ such that $\widehat\xi_\kappa(\mu_j)=j/M$,
$j=1,\dots,N$. Their explicit representations are
 \begin{equation}\label{PPkk}
 \wh{P}_k= \f{ 1 }{\wh{\rho}_{\kappa}(\mu_{p_k})}
 \prod_{j=1}^N \paa{ \f{j-p_k-\wh{F}(\mu_{p_k})}{j-p_k-\wh{F}(\lambda_{j})}
  \cdot\f{ \wh{\vp}_{\kappa}\pa{\mu_{p_k},\mu_j}  }{  \wh{\vp}_{\kappa}\pa{\mu_{p_k},\la_j} }  }^{\! 2}
 \Gamma^2\!\left(
 \begin{array}{c}
 \! p_k \ ,\  p_k-N+ \wh{F}(\mu_{p_k}) \! \\
 \! p_k+ \wh{F}(\mu_{p_k})\  ,\ p_k-N \!
 \end{array}\right)  .
 \end{equation}
The representations for $\wh{H}_k$ are similar to \eqref{PPkk}, but
they depend on $\mu_{h_k}$ and $h_k$:
 \begin{multline}
\wh{H}_k= \f{ \sin^2(\pi \wh{F} (\mu_{h_k})) }{ \pi^2
\wh{\rho}_\kappa\pa{\mu_{h_k}}} \,
\prod_{j=1}^N\paa{\frac{j-h_k-\wh{F} (\lambda_{j})}{j-h_k-\wh{F}
(\mu_{h_k})}
      \cdot\f{ \wh{\vp}_{\kappa}\pa{\mu_{h_k},\la_j}  }{  \wh{\vp}_{\kappa}\pa{\mu_{h_k},\mu_j} }  }^{\! 2} \\
 \times \Gamma^2\! \left(
 \begin{array}{c}
 \! h_k+ \wh{F} (\mu_{h_k})\ ,\ N+1-h_k- \wh{F} (\mu_{h_k}) \! \\
 \! h_k\  ,\ N+1-h_k \!
 \end{array}\right) .
 \end{multline}
Finally, $D_{N,0}$ does not depend on the  particle/hole rapidities
and the corresponding integers
 \begin{multline}\label{D0}
 D_{N,0}=\prod_{j=1}^{N}  \frac{\wh\rho_\kappa(\lambda_{j})}
 {\wh\rho(\lambda_{j})}
 \prod_{j,k=1}^{N} \frac{\wh \varphi_\kappa(\lambda_{j},\lambda_{k}) \, \wh \varphi_\kappa(\mu_{j},\mu_{k})}
 {\wh \varphi^{\,2}_\kappa(\mu_{j},\lambda_{k})}\\
 \times
 \prod\limits_{j>k}^N \left(1-\frac{\wh F(\lambda_j)-\wh  F(\lambda_k)}{j-k}\right)^2
 \prod\limits_{j,k=1\atop{j\ne k}}^N \left(1-\frac{\wh  F(\lambda_j)}{j-k}\right)^{-2}
 \prod_{j=1}^N \paa{ \frac{\sin\bigl(\pi \wh  F(\lambda_j)\bigr)}{\pi \wh F(\lambda_j)} }^2.
 \end{multline}

The fact that $\wh{E}_0 \to E_0$ follows from $\wh{\vp}_{\kappa} \to
\vp$. The calculation of the limit of the first product in
\eqref{D0} is based on the definition of the shift function
\eqref{SP-shiftF-mod}:
\begin{multline}\label{Singl-prod}
  \prod_{j=1}^N\frac{\widehat\rho_\kappa(\lambda_j)}{\widehat\rho(\lambda_j)}
 =
 \prod_{j=1}^N\left(1+\frac{(\widehat\xi'_\kappa(\lambda_j)-\widehat\xi'(\lambda_j))}{\widehat\rho(\lambda_j)}\right)\\
 = \prod_{j=1}^N\left(1-\frac{\widehat
 F'(\lambda_j)}{M\widehat\rho(\lambda_j)}\right)
 \to \exp\left(-\int\limits_{-q}^q
 F'(\lambda)\,d\lambda\right)=e^{F_- - F_+}.
  \end{multline}
The limit of the remaining part of $D_{N,0}$ was computed in
\cite{KitKMST09c}, what gives us
% Here, we only remind the result:
%
\begin{equation}
 D_{N,0}=M^{-F_+^2-F_-^2} \cdot \mc{D}^{\pa{z}}[F] \cdot \pa{1+\e{O}\pa{\f{\log M}{M}}}.
 \end{equation}
Finally, the limits $\wh{P}_k\to P_N(\mu_{p_k},p_k)$ and
$\wh{H}_k\to H_N(\mu_{h_k},h_k)$ follow from
\begin{equation}
 \lim_{N,M\to\infty} \prod_{j=1}^N \paa{ \f{j-k-\wh{F}(\mu_{k})}{j-k-\wh{F}(\lambda_{j})}
  \f{ \wh{\vp}_{\kappa}\pa{\mu_{k},\mu_j}  }{  \wh{\vp}_{\kappa}\pa{\mu_{k},\la_j} }  }^2 = e^{J
  [F]\pa{\mu_k}} \; .
\end{equation}
This formula is proved in appendix~\ref{Sproof} (see
\eqref{prod-sing}). \qed

Assuming general positions of the particles and holes, we can not
make further simplifications in the formula for $D_{N}^{\pa{z}}$.
Therefore we now study two limiting cases of interest, exactly as we
did for the smooth parts: first when all particles and holes are
separated from the Fermi boundaries, and then when all particles and
holes are in the vicinity of the Fermi zone (critical form factor in
the class $\mathbf{P}_{r}$).

\begin{cor}
Suppose that, in the thermodynamic limit, particles and holes are
separated from the Fermi boundaries: $\mu_{p_a} \not=\pm q$ and $\mu_{h_a}\not= \pm q$.
Then the thermodynamic limit of the factor $D_N^{\pa{z}}$ becomes a smooth function of the rapidities
$\{\mu_{p}\}$ and $\{\mu_{h}\}$:
 \begin{equation}\label{def-Dn}
D_{N}^{(z)} = M^{-2n-F_-^2-F_+^2} \cdot
D_{n}^{(z)}(\{\mu_p\},\{\mu_h\})
                         \cdot \pa{1+\e{O}\paf{\log M}{M}},
 \end{equation}
where $D_{n}^{(z)}(\{\mu_p\},\{\mu_h\})$ is given by the expression \eqref{lim-Dz} for $\mathcal{D}_{zz}$.
\end{cor}

\proof If the rapidities of particles and holes are separated from
the Fermi boundaries, then all the arguments of the
$\Gamma$-functions in \eqref{Ekp} and $\eqref{Ekh}$ are large and we
can apply the Stirling formula for their simplification. Then
 \begin{multline}\label{lim-Gam}
 \lim_{N,M\to\infty}\Gamma^2\left(
 \begin{array}{c}
 p_k\  ,\ p_k-N+ \widehat F(\mu_{p_k})\  , \ N+1-h_k-
 \widehat F(\mu_{h_k})\  ,\ h_k+ \widehat F(\mu_{h_k})\\
 p_k-N\  ,\ p_k+ \widehat F(\mu_{p_k})\  , \ N+1-h_k\  ,\ h_k
 \end{array}\right)\\
 =\lim_{N,M\to\infty}\left(\frac{p_k-N}{p_k}\right)^{2F(\mu_{p_k})}
 \left(\frac{N-h_k}{h_k}\right)^{-2F(\mu_{h_k})}\\
 = \left(\frac{p(\mu_{p_k})-p(q)}{p(\mu_{p_k})-p(-q)}\right)^{2F(\mu_{p_k})}
 \left(\frac{p(q)-p(\mu_{h_k})}{p(\mu_{h_k})-p(-q)}\right)^{-2F(\mu_{h_k})},
 \end{multline}
where we have used \eqref{0-Dmom}. Combining \eqref{lim-Gam} with
the explicit form \eqref{fonction J} of $J\pa{\om}$, we arrive at
 \begin{multline}\label{lim-Ek}
 P_N\pa{\mu_{p_k},p_k}\,H_N\pa{\mu_{h_k},h_k}=\frac{\sin^2(\pi F(\mu_{h_k}))}{\pi^2
\rho(\mu_{h_k})\rho(\mu_{p_k})} \\
 \times\exp\! \left\{2\, \int\limits_{-q}^{q}\hspace{-4.2mm}\diagup\hspace{.5mm}
 F(\lambda)\bigl(\coth(\lambda-\mu_{p_k})-\coth(\lambda-\mu_{h_k})\bigr)
 \,d\lambda\right\}.
 \end{multline}

The limit of $\widehat E_0$ is almost trivial:
 \begin{equation}\label{lim-E0}
 \lim_{N,M\to\infty}\widehat E_0M^{2n}=\left(\det_n\frac1{\sinh(\mu_{p_j}
 -\mu_{h_k})}\right)^2.
 \end{equation}
Gathering all these results, we get the claim. \qed

\medskip

Hence, in the limit when the rapidities of particles and holes
remain at finite distance from the Fermi boundaries, the discrete
part $D_{N}^{(z)}$, up to an $M$-dependent normalization, can  be
approximated by a smooth function of these rapidities. However, as
soon as some rapidities go to $\pm q$, then certain integrals in the
second line of \eqref{lim-Dz} become ill-defined. There can also
appear singularities in the Cauchy determinants present in the first
line of \eqref{lim-Dz}. Therefore, the discrete structure of
$D_{N}^{(z)}$ manifests itself when particles and holes condensate
on the Fermi boundaries.

We now describe the asymptotic behavior of an excited state
$|\psi_\kappa(\{\mu_{\ell_j}\}\rangle$ belonging to the
$\mathbf{P}_{r}$ class. This means that the rapidities of all
particles and holes are located in vicinities of the Fermi
boundaries with the condition \eqref{Nak-mom}. Any such state, can
be parameterized by a set of integers as given in
\eqref{spec-p}.

\begin{cor}
The asymptotic behavior of  $D_N^{\pa{z}}$  for an excited state in
the $\mathbf{P}_{r}$ class parameterized as in \eqref{spec-p} reads
 \begin{equation}\label{D-fin}
 D_N^{(z)}(\{+q\}_{n_p^+}\cup\{-q\}_{n_p^-},
 \{+q\}_{n_h^+}\cup\{-q\}_{n_h^-}) =M^{-F_{r,+}^2 -F_{r,-}^2} \, D_{n,r}^{(z)} \,
   \pa{1+\e{O}\paf{\log M}{M} } ,
 \end{equation}
where $D_{n,r}^{(z)}$ is given by the expression \eqref{D-fin-z} for $\mathcal{D}_{zz}$.
\end{cor}

\begin{rem}
Thence, when particles and holes are in the vicinity of the Fermi boundaries, the behavior of
$D_N^{\pa{z}}$ cannot be called a thermodynamic limit. Indeed, the factors $R_{n_p^+,n_h^+}$ and $R_{n_p^-,n_h^-}$,
and thus the form factor, depend explicitly on the microscopic characteristics
$p^\pm$ and $h^\pm$ of the excited state.
\end{rem}

\proof In this case, the limits $E_0$, $P_{N}(\mu_{p_k},p_k)$ and
$H_{N}(\mu_{h_k},h_k)$ should be re-calculated.

Let us first consider the coefficient $\widehat E_0$. Using that
$\varphi(q,q)=\varphi(-q,-q)=\rho^{-1}(q)$ and
$\varphi(q,-q)=\varphi(-q,q)=\sinh(2q)/D$, we get
 \begin{equation}\label{lim-E0-qq}
 \lim_{N,M\to\infty}\widehat E_0=\left(\frac
 D{\rho(q)\sinh(2q)}\right)^{2r^2}\rho^{2n}(q)\lim_{N,M\to\infty}\left(
 \det_n\frac1{p_j-h_k}\right)^2.
 \end{equation}
Using the specific parametrization \eqref{spec-p} of the integers
belonging to the $\mathbf{P}_{r}$ class we see that the
thermodynamic limit of the Cauchy determinant in \eqref{lim-E0-qq}
can be factorized in two parts:
 \begin{equation}\label{lim-Cauchy-int}
 \left(\det_n\frac1{p_j-h_k}\right)^2\to N^{-2r^2} \, L_{n_p^+,n_h^+}(\{p^+\},\{h^+\}) \;
 L_{n_p^-,n_h^-}(\{p^-\},\{h^-\}),
 \end{equation}
with
 \begin{equation}\label{def-L}
 L_{n,m}(\{p\},\{h\})=\frac{\prod\limits_{j>k}^n(p_j-p_k)^2\prod\limits_{j>k}^m(h_j-h_k)^2}
 {\prod\limits_{j=1}^n\prod\limits_{k=1}^m(p_j+h_k-1)^2} \; .
 \end{equation}
This leads eventually to the following estimate
 \begin{equation}\label{lim-E0-qq-1}
 \lim_{N,M\to\infty}\widehat E_0\left(M\rho(q)\sinh(2q)\right)^{2r^2}
 =\rho^{2n}(q) \, L_{n_p^+,n_h^+}(\{p^+\},\{h^+\}) \;
 L_{n_p^-,n_h^-}(\{p^-\},\{h^-\}).
 \end{equation}

When computing the limits of $\widehat P_{k}$ and $\widehat H_{k}$,
some of the arguments of the $\Gamma$-functions are not large
anymore. Therefore, we can only use Stirling formula partly. We have
 \begin{multline}\label{prod-Ek-qq}
 \prod_{k=1}^n\Gamma^2\left(
 \begin{array}{c}
 p_k\  ,\ p_k-N+\widehat F\pa{\mu_{p_k}} \  , \ N+1-h_k-\widehat F\pa{\mu_{h_k}} \
 ,\ h_k+\widehat F\pa{\mu_{h_k}}\\
 p_k-N\  ,\ p_k+\widehat F\pa{\mu_{p_k}}\  , \ N+1-h_k \   ,\ h_k
 \end{array}\right)\\
 \sim N^{-2r(F_+ + F_-)}\;
 \Gamma^2
 \left(
 \begin{array}{c}
 \{p^++F_+\}\  ,\ \{h^+-F_+\}\  ,\ \{h^-+F_-\} \  ,\ \{p^--F_-\}\\
 \{p^+\}\  ,\ \{h^+\}\  ,\ \{h^-\}\  , \  \{p^-\}
 \end{array}\right).
 \end{multline}
We precise that the notation $\{p^+ + F_+\}$ means $  \{p^+_j +
F_+\}_{j=1}^{n^+_p}$ and similarly for other sets of parameters in
\eqref{prod-Ek-qq}. Combining all these results we obtain
 \begin{multline}\label{Ek-qq}
 \lim_{N,M\to\infty}\widehat E_0 \left(M\rho(q)\sinh(2q)\right)^{2r^2+2r(F_++F_-)}
 \prod_{k=1}^n \widehat P_k \widehat  H_k =
 \\
 \left(\frac{\sin(\pi F_+)}{\pi}\right)^{2n^+_h}
 \left(\frac{\sin(\pi F_-)}{\pi}\right)^{2n^-_h}
 R_{n_p^+,n_h^+}(\{p^+\},\{h^+\}|F_+) \; R_{n_p^-,n_h^-}(\{p^-\},\{h^-\}|-F_-)\\
 \times
  \exp\left\{2r
 \int\limits_{-q}^{q}\frac{F_+-F(\lambda)}{\tanh(q-\lambda)}\,d\lambda+
 2r
 \int\limits_{-q}^{q}\frac{F_--F(\lambda)}{\tanh(q+\lambda)}\,d\lambda\right\}.
 \end{multline}
After some simple algebra we get the claim. \qed

To conclude this section we would like to stress that the evaluation
of the $S_N^z$ thermodynamic limit was done without using  the
explicit form  of the shift function $F(\lambda)$
\eqref{F-solution}. We have used this representation only in
\eqref{Fr-Z} in order to relate the smooth part of the critical form
factors of $\mathbf{P}_r$ class with the one corresponding to the
$(\alpha+r)$-twisted ground state. However in all other respects
$F(\lambda)$ played the role of a free functional parameter.

%%%%%%%%%%%%%%%%%%%%%%%%%%%%%%%%%%%%%%%%%%%%%%%%%%%%%%%%%%%%%%%%%%%%%%%%%%%%%%%%%%%%%%%%
%%%%%%%%%%%%%%%%%%%%%%%%%%%%%%%%%%%%%%%%%%%%%%%%%%%%%%%%%%%%%%%%%%%%%%%%%%%%%%%%%%%%%%%%
%%%%%%%%%%%%%%%%%%%%%%%%%%%%%%%%%%%%%%%%%%%%%%%%%%%%%%%%%%%%%%%%%%%%%%%%%%%%%%%%%%%%%%%%
%%%%%%%%%%%%%%%%%%%%%%%%%%%%%%%%%%%%%%%%%%%%%%%%%%%%%%%%%%%%%%%%%%%%%%%%%%%%%%%%%%%%%%%%
%%%%%%%%%%%%%%%%%%%%%%%%%%%%%%%%%%%%%%%%%%%%%%%%%%%%%%%%%%%%%%%%%%%%%%%%%%%%%%%%%%%%%%%%
%%%%%%%%%%%%%%%%%%%%%%%%%%%%%%%%%%%%%%%%%%%%%%%%%%%%%%%%%%%%%%%%%%%%%%%%%%%%%%%%%%%%%%%%

\section{The scalar product $S_N^{+}$\label{WWS+1/2}}

We now study the scalar product $S_N^{+}$   for
$\ket{\psi_\kappa(\{\mu\})}$ being an $\alpha$-twisted excited state
with $n$ particles and $n$ holes in the $N+1$ sector ({\it i.e.}
$N_\kappa=N+1$). As previously, $\{\mu_{p_a}\}_{a=1}^n$ and
$\{\mu_{h_a}\}_{a=1}^n$ denote the rapidities of the particles and
holes respectively. The notation $\widehat F(\lambda)$ now means the
shift function in the $N+1$ sector, whose thermodynamic limit
$F^{(+)}(\lambda)$ is given by \eqref{F-solution+}. However, like in
the previous section, we omit the superscript $(+)$ in order to
lighten the notations.

\subsection{Representation for the scalar product $S_N^+$}

In comparison with the $S_N^{z}$ case, the excited state now
depends on $N+1$ parameters $\mu$.  Using the determinant representation given in
Proposition \ref{proposition produits scalaire Fredholm} as well as the norm formula of Proposition \ref{proposition norme},
we get the following representation for $S_N^{+}$ \eqref{SP-ff}:
 \begin{equation}\label{SN+1/2-prom}
 S_N^{+}(\{\mu_p\},\{\mu_h\})=
 {\cal A}^{(+)}_{N}(\{\mu_p\},\{\mu_h\})
 \cdot \exp \paa{C^{(+)}_{N}(\{\mu_p\},\{\mu_h\})} \cdot D^{(+)}_{N}(\{\mu_p\},\{\mu_h\}).
 \end{equation}
Here
 \begin{equation}\label{Dn+1/2}
 D^{(+)}_{N}(\{\mu_p\},\{\mu_h\})= \frac{ M\pi^2 \wh{\rho}\pa{\la_{N+1}} }{ \sin^2 \pi \wh{F}\pa{\la_{N+1}} }
  \frac{ \pl{a=1}{N+1}\sinh^2(\la_{N+1}-\mu_{\ell_a})}
                                        {\pl{a=1}{N}\sinh^2(\la_{N+1}-\la_a)}
 \cdot D^{(z)}_{N+1} (\{\mu_p\},\{\mu_h\})\, ,
 \end{equation}
where $D^{(z)}_{N+1}$ is given by \eqref{6-DNk} with $N$ replaced by
$N+1$. It is expressed in terms of the $N+1$ parameters
$\mu_{\ell_a}$ as well as the $N+1$ parameters $\la_j$ defined by
$\wh{\xi}\pa{\la_j}=\tf{j}{M}$, $\wh{\xi}(\lambda)$ being the
counting function \eqref{BE-cf} for the ground state. In other words
$\la_j$, $j=1,\dots, N$, are Bethe roots for the ground state,
whereas $\la_{N+1}$ is the point where $M\wh{\xi}(\lambda)$ takes
its next integer value, \textit{ie} $M\wh{\xi}\pa{\la_{N+1}}=N+1$.

The coefficient $C^{(+)}_{N}$ is now modified according to
 \begin{equation}\label{Cn+1/2}
 \exp\pac{  C^{(+)}_{N}(\{\mu_p\},\{\mu_h\})}
 = \prod\limits_{a=1}^{N+1}\left|\frac{\sinh(\la_{a}-\lambda_{N+1}-i\zeta)}
                                     {\sinh(\mu_{\ell_{a}}-\la_{N+1}-i\zeta)}\right|^2
  \; \cdot \; \exp\pac{ C_{N+1}^{(z)}(\{\mu_p\},\{\mu_h\}) },
 \end{equation}
where, up to the evident modifications stemming from $N\to N+1$,
$C_{N+1}^{\pa{z}}$ is given by \eqref{W}. Finally the factor ${\cal
A}^{(+)}_{N}$ has the following form:
 \begin{equation}\label{An+1/2}
 {\cal A}^{(+)}_{N}(\{\mu_p\},\{\mu_h\})=\frac{\sin\zeta}{2\pi \kappa}
 \left|\frac{\prod\limits_{a=1}^{N}\sinh(\lambda_a-\frac{i\zeta}2)}
             {\prod\limits_{a=1}^{N+1}\sinh(\mu_{\ell_a}-\frac{i\zeta}2)}\right|^2\cdot
 \frac{\left|\det\left[ I+\frac1{2\pi i}\wh
                         U^{(+)}(w,w')\right] \right|^2}
        {\det_{N+1} \Theta_{jk}^{(\mu)} \cdot \det_N \Theta_{jk}^{(\lambda)}},
 \end{equation}
and we refer to \eqref{2-U2} for the definition of the kernel
$\wh{U}^{(+)}(w,w')$.

\subsection{Thermodynamic limit of the smooth part}

The smooth part of $S_N^{+}$ consists of the coefficients ${\cal
A}^{(+)}_{N}$ and $\exp C^{(+)}_{N}$. The computation of their
thermodynamic limits $C_{n}^{(+)}$ and ${\cal A}_n^{(+)}$ does not
contain any subtleties comparing with the derivation described in
section~\ref{TL-SP-z}. The use of the modified shift function
\eqref{F-mod}  formally reduces the computations to the case where
the excited state does not contain particles and holes. In this way,
starting from the representations \eqref{Cn+1/2}, \eqref{An+1/2} we
arrive at the equations \eqref{Cn+}, \eqref{An+}. One can easily see
that the coefficients $C_{n}^{(+)}$ and ${\cal A}_n^{(+)}$ are well
defined for general positions of particles and holes. In particular,
for the critical form factors of the $\mathbf{P}_r$ class the smooth
part effectively depends on the shift function
$F_r(\lambda)=F(\lambda)+r$.

\subsection{Thermodynamic limit of the discrete part}

\begin{prop}
The discrete part $D_N^{(+)}$ behaves in the thermodynamic as
 \begin{multline}\label{DNk-new3}
 D_N^{(+)}(\{\mu_p\},\{\mu_h\})= M^{-F_-^2-(F_+ + 1)^2}
E_0\pa{\paa{\mu_h, \mu_p},\paa{h_a, p_a }}
 \cdot \prod_{k=1}^{n} \widetilde H_{N+1}\pa{\mu_{h_k},h_k}  \\
 \times  \prod_{k=1}^{n} \widetilde P_{N+1}\pa{\mu_{p_k},p_k}
 \cdot
 \mc{D}^{\pa{+}}[F] \cdot \pa{1+\e{O}\paf{\log M}{M}}  ,
 \end{multline}
where
 \begin{align}\label{til-HP}
 \widetilde
 H_{N+1}\pa{\mu_{h_k},h_k}&=H_{N+1}\pa{\mu_{p_k},p_k}(N+1-h_k-F(\mu_{h_k}))^{-2}\varphi^{-2}(q,\mu_{h_k}),\\
 \widetilde
 P_{N+1}\pa{\mu_{p_k},p_k}&=P_{N+1}\pa{\mu_{p_k},p_k}(N+1-p_k-F(\mu_{p_k}))^2\varphi^2(q,\mu_{p_k}).
 \end{align}
The functions entering these definitions are the same as in
Proposition~\ref{Proposition asymptotique Dz}, but  one should
replace  $N$ by $N+1$ for the coefficients $H_{N+1}$ and $P_{N+1}$.
We also remind that the functional $\mc{D}^{\pa{+}}$ is given in
\eqref{fonctionnelle D+}.
\end{prop}

\proof

As the limit of $D_{N+1}^{(z)}$ is already known, it remains to evaluate the limit of the product
 \begin{equation}\label{P}
 P=\frac{ \pl{a=1}{N+1}\sinh^2(\la_{N+1}-\mu_{\ell_a})}
                                        {\pl{a=1}{N}\sinh^2(\la_{N+1}-\la_a)}\, .
 \end{equation}
Harping on the steps in the analysis of $D_{N}^{(z)}$, we multiply
and divide by the counting function $\wh{\xi}(\lambda)$
\eqref{BE-cf}. We obtain
 \begin{equation}\label{P-xi}
 P=  \f{ \pl{a=1}{N+1}  \pa{N+1-\ell_a-\wh{F}\pa{\mu_{\ell_a}} }^2 }{ M^2 \pl{a=1}{N} \pa{N+1-a}^2 }
 \cdot \f{ \pl{a=1}{N+1} \wh{\vp}^2\pa{\la_{N+1},\mu_{\ell_a}} }{ \pl{a=1}{N} \wh{\vp}^2\pa{\la_{N+1},\la_{a}}}\,
 .
 \end{equation}
Then using Corollary~\ref{sing-product} we find that, in the
thermodynamic limit, the product $P$ behaves as
 \begin{equation}\label{limP-gen}
 P \to   \f{ e^{J[F]\pa{q}} }{M^2 \rho^2\pa{q} } \;
 \Gamma^2\left(\begin{array}{c}
  N+1-F_+\\
  N+1 \  , \  -F_+ \end{array}\right)  \cdot  \pl{k=1}{n} \left[\frac{N+1-p_k-F(\mu_{p_k})}
  {N+1-h_k-F(\mu_{h_k})}\cdot\frac{\varphi(q,\mu_{p_k}) }{ \varphi(q,\mu_{h_k}) }
  \right]^2
  ,
 \end{equation}
with $J[F]\pa{q}$ defined as in \eqref{fonction J}. Using now that
$N$ is large, we have
\begin{equation}
e^{J[F]\pa{q}}
            \Gamma^2 \! \left(\begin{array}{c}  N+1-F_+\\
                       \!   N+1\ , \  -F_+ \! \end{array}\right)
 \to  \f{\Ga^2(1+F_+)\sin^2\pi F_+}{ \pi^2 \pa{\rho\pa{q} \s{2q}M
}^{2F_+} }\, \exp\paa{2\Int{-q}{q}
\f{F\pa{\la}-F_+}{\tanh\pa{\la-q}} }.
\end{equation}
 Taking into account \eqref{fonctionnelle Dz}, \eqref{fonctionnelle D+}
 we find
\begin{equation}\label{limit-P}
 \mc{D}^{\pa{z}}[F]\cdot P\cdot\frac{M\pi^2\widehat\rho(\lambda_{N+1})}
 {\sin^2(\pi\widehat F(\lambda_{N+1}))}\to
 \frac{\mc{D}^{\pa{+}}[F]}{ M^{2F_++1} }
 \pl{k=1}{n} \left[\frac{N+1-p_k-F(\mu_{p_k})}
  {N+1-h_k-F(\mu_{h_k})}\cdot\frac{\varphi(q,\mu_{p_k}) }{ \varphi(q,\mu_{h_k}) }
  \right]^2.
\end{equation}
It remains to combine the obtained result with the known limit of
$D_{N+1}^{(z)}$ and to substitute it into \eqref{Dn+1/2}. A few
algebraic manipulations lead then to the claim. \qed

We now particularize this result to the two limiting  cases we
considered previously, namely when all particles and holes are
separated from the Fermi boundaries, or when all particles and holes
are on the Fermi boundaries.

\begin{cor}
Suppose that, in the thermodynamic limit, particles and holes are
separated from the Fermi boundaries: $\mu_{p_a} \not=\pm q$ and $\mu_{h_a}\not= \pm q$.
Then the thermodynamic limit of $D_N^{\pa{+}}$ becomes a smooth function of the rapidities
$\{\mu_{p}\}$ and $\{\mu_{h}\}$:
 \begin{equation}\label{def-Dn-bis}
D_{N}^{(+)} = M^{-2n-F_-^2-(F_++1)^2} \cdot
D_{n}^{(+)}(\{\mu_p\},\{\mu_h\})
                        \cdot \pa{1+\e{O}\paf{\log M}{M}},
 \end{equation}
where $D_{n}^{(+)}(\{\mu_p\},\{\mu_h\})$ is given by the expression \eqref{D-separ} for $\mathcal{D}_{+-}$.
\end{cor}

\proof

If the rapidities of particles and holes are separated from the
Fermi boundaries, then
 \begin{equation}\label{limP-sep}
\pl{k=1}{n} \left[\frac{N+1-p_k-F(\mu_{p_k})}
  {N+1-h_k-F(\mu_{h_k})}\cdot\frac{\varphi(q,\mu_{p_k}) }{ \varphi(q,\mu_{h_k}) }
  \right]^2 \to \pl{k=1}{n} \pa{\f{\s{q-\mu_{p_k}} }{ \s{q-\mu_{h_k}} }
  }^2.
 \end{equation}
Substituting this limit into \eqref{DNk-new3} and using the results
of the previous section we arrive at the statement. \qed

\begin{cor}
The asymptotic behavior of  $D_N^{\pa{+}}$  for an excited state in
the $\mathbf{P}_{r}$ class parameterized as in \eqref{spec-p} reads
 \begin{equation}\label{D-fin-bis}
 D_N^{(+)}(\{+q\}_{n_p^+}\cup\{-q\}_{n_p^-},
 \{+q\}_{n_h^+}\cup\{-q\}_{n_h^-}) =M^{-(F_{r,+}+1)^2 -F_{r,-}^2}\,  D_{n,r}^{(+)} \, \pa{1+\e{O}\paf{\log M}{M} } ,
 \end{equation}
where $D_{n,r}^{(+)}$ is given by the expression \eqref{D-fin+-} for $\mathcal{D}_{+-}$.
\end{cor}

\proof

In this case
 \begin{equation}\label{limP-col}
\pl{k=1}{n} \left[\frac{N+1-p_k-F(\mu_{p_k})}
  {N+1-h_k-F(\mu_{h_k})}\cdot\frac{\varphi(q,\mu_{p_k}) }{ \varphi(q,\mu_{h_k}) }
  \right]^2 \to \bigl(M\rho(q)\sinh(2q)\bigr)^{-2r} \frac{\prod_{k=1}^{n_p^+}(p_k^++F_+)^2}
  {\prod_{k=1}^{n_h^+}(h_k^+-1-F_+)^2}.
 \end{equation}
Substituting this limit into \eqref{DNk-new3} and using the results
of the previous section we arrive at the statement. \qed

%%%%%%%%%%%%%%%%%%%%%%%%%%%%%%%%%%%%%%%%%%%%%%%%%%%%%%%%%%%%%%%%%%%%%%%%%%%%%%%%%%%%%%%%
%%%%%%%%%%%%%%%%%%%%%%%%%%%%%%%%%%%%%%%%%%%%%%%%%%%%%%%%%%%%%%%%%%%%%%%%%%%%%%%%%%%%%%%%
%%%%%%%%%%%%%%%%%%%%%%%%%%%%%%%%%%%%%%%%%%%%%%%%%%%%%%%%%%%%%%%%%%%%%%%%%%%%%%%%%%%%%%%%
%%%%%%%%%%%%%%%%%%%%%%%%%%%%%%%%%%%%%%%%%%%%%%%%%%%%%%%%%%%%%%%%%%%%%%%%%%%%%%%%%%%%%%%%
%%%%%%%%%%%%%%%%%%%%%%%%%%%%%%%%%%%%%%%%%%%%%%%%%%%%%%%%%%%%%%%%%%%%%%%%%%%%%%%%%%%%%%%%
%%%%%%%%%%%%%%%%%%%%%%%%%%%%%%%%%%%%%%%%%%%%%%%%%%%%%%%%%%%%%%%%%%%%%%%%%%%%%%%%%%%%%%%%

\section*{Conclusion}

In this article we have provided the first steps towards  the study of correlation
functions \textit{via} the form factor approach. For this purpose we
have calculated the thermodynamic limits of the so-called
particle/hole form factors. Although we have considered the specific
case of the XXZ chain, our method is straightforwardly  applicable
to other massless integrable models solvable by means of the
algebraic Bethe Ansatz and having determinant representations of their form factors. In particular, one can use it for the
calculation of the thermodynamic limit of form factors in the model
of one-dimensional bosons. Determinant representations for the form factors of this model in the finite volume were
given in \cite{KojKS1997,KorS1999}.  Their thermodynamic limit can be directly obtained from the results given  in the present
article.

%\noindent
Our results show that the idea to replace, in the thermodynamic
limit, the sum over excited states by an integration with respect to
particles and holes should be essentially modified. In particular, a
formal replacement of the discrete sums over $n$-particle/hole
excited states $|\psi'(\{\mu_p\},\{\mu_h\})\rangle$ by integrals as
 \begin{equation}\label{form-replac}
\sum_{|\psi'(\{\mu_p\},\{\mu_h\})\rangle} {\cal
F}^{(s)}_{\psi_g\,\psi'}(m)\; {\cal F}^{(s')}_{\psi'\,\psi_g}(m')
\to M^{2n}\int {\cal F}^{(s)}_{\psi_g\,\psi'}(m)\; {\cal
F}^{(s')}_{\psi'\,\psi_g}(m')\;\prod_{j=1}^n\rho(\mu_{p_j})\rho(\mu_{h_j})
d\mu_{p_j}d\mu_{h_j},
 \end{equation}
leads to senseless results. First of all, even in the region where
the form factors depend smoothly on $\{\mu_p\}$ and $\{\mu_h\}$, the
coefficient $M^{2n}$ does not compensate the factors
$M^{-\theta_{ss'}}$, therefore in the above expression, the
\textit{r.h.s.} vanishes. On the other hand, the corresponding
integral becomes divergent when the rapidities of particles and
holes approach the Fermi boundaries (see \eqref{lim-Dz},
\eqref{D-separ}). Lastly, due to the discrete structure of the form
factors when some rapidities agglomerate on the Fermi boundaries, one
cannot replace the sum over the excited states in the
$\mathbf{P}_{r}$ class by an integral.  For such excited states, one
has to take the microscopic structure of the excited state into
account and perform the discrete sums over the parameters $p^{\pm}$,
$h^{\pm}$ \eqref{spec-p}.

%\noindent
In a forthcoming publication \cite{KitKMST10b}, we will describe a
way to overcome these difficulties. We will show that in the
asymptotic regime (large lattice distances $m$ between local spin
opertors) only particles and holes having their rapidities close to
the Fermi boundaries contribute to the form factor sums. In such a
limit, one can send the rapidities of the particles and holes to
$\pm q$ in the smooth part of the form factor. Hence, it becomes a
constant that only depends on the $\mathbf{P}_{r}$ class ${\cal
S}_{ss'}$=${\cal S}_{ss'}^{(r)}(\{\pm q\},\{\pm q\})$. In its turn
the summation over the integers $\{p\}$ and $\{h\}$ present in the
discrete part of the form factors leads to the natural re-scaling
$M\to 2\pi m$ of the system size $M$ into the distance $m$ between
the operators. This mechanism explains the appearance of the
critical exponents $\theta_{ss'}$ in the asymptotic behavior of the
correlation functions.

%%%%%%%%%%%%%%%%%%%%%%%%%%%%%%%%%%%%%%%%%%%%%%%%%%%%%%%%%%%%%%%%%%%%%%%%%%%%%%%%%%%%%%%%
%%%%%%%%%%%%%%%%%%%%%%%%%%%%%%%%%%%%%%%%%%%%%%%%%%%%%%%%%%%%%%%%%%%%%%%%%%%%%%%%%%%%%%%%
%%%%%%%%%%%%%%%%%%%%%%%%%%%%%%%%%%%%%%%%%%%%%%%%%%%%%%%%%%%%%%%%%%%%%%%%%%%%%%%%%%%%%%%%
%%%%%%%%%%%%%%%%%%%%%%%%%%%%%%%%%%%%%%%%%%%%%%%%%%%%%%%%%%%%%%%%%%%%%%%%%%%%%%%%%%%%%%%%
%%%%%%%%%%%%%%%%%%%%%%%%%%%%%%%%%%%%%%%%%%%%%%%%%%%%%%%%%%%%%%%%%%%%%%%%%%%%%%%%%%%%%%%%
%%%%%%%%%%%%%%%%%%%%%%%%%%%%%%%%%%%%%%%%%%%%%%%%%%%%%%%%%%%%%%%%%%%%%%%%%%%%%%%%%%%%%%%%

\section*{Acknowledgements}

J. M. M., N. S. and V. T. are supported by CNRS.  We also
acknowledge  the support from the GDRI-471 of CNRS "French-Russian
network in Theoretical and Mathematical  Physics" and
RFBR-CNRS-09-01-93106L-a. N. K., J. M. M. and V. T are also supported by the ANR grant DIADEMS 10 BLAN 012004 and  N. S.  by the Program of
RAS Mathematical Methods of the Nonlinear Dynamics,
RFBR-11-01-00440-a. K. K. K. is supported by the EU Marie-Curie
Excellence Grant MEXT-CT-2006-042695. N. K., N. S. and K. K. K would
like to thank the Theoretical Physics group of the Laboratory of
Physics at ENS Lyon for hospitality, which makes this collaboration
possible. N.K. and V. T. would  like to thank LPTHE (Paris VI University) for
hospitality.

%%%%%%%%%%%%%%%%%%%%%%%%%%%%%%%%%%%%%%%%%%%%%%%%%%%%%%%%%%%%%%%%%%%%%%%%%%%%%%%%%%%%%%%%
%%%%%%%%%%%%%%%%%%%%%%%%%%%%%%%%%%%%%%%%%%%%%%%%%%%%%%%%%%%%%%%%%%%%%%%%%%%%%%%%%%%%%%%%
%%%%%%%%%%%%%%%%%%%%%%%%%%%%%%%%%%%%%%%%%%%%%%%%%%%%%%%%%%%%%%%%%%%%%%%%%%%%%%%%%%%%%%%%
%%%%%%%%%%%%%%%%%%%%%%%%%%%%%%%%%%%%%%%%%%%%%%%%%%%%%%%%%%%%%%%%%%%%%%%%%%%%%%%%%%%%%%%%
%%%%%%%%%%%%%%%%%%%%%%%%%%%%%%%%%%%%%%%%%%%%%%%%%%%%%%%%%%%%%%%%%%%%%%%%%%%%%%%%%%%%%%%%
%%%%%%%%%%%%%%%%%%%%%%%%%%%%%%%%%%%%%%%%%%%%%%%%%%%%%%%%%%%%%%%%%%%%%%%%%%%%%%%%%%%%%%%%

\appendix

\section{Summation identities\label{Sproof}}

\begin{lemma}\label{m>1}
Let $f\in C^{1} ( [0,a] )$ for some $a>D$. Let $(h_M)$ be a sequence of integers such that $\frac{h_M}{M}$ tends to some finite value $x_h\in [0,a]$ when $M\to +\infty$.
Then, in the limit $N,M \to +\infty$,
$\tf{N}{M}\to D$, the following sums vanish:
 \begin{equation}
    \lim_{N,M\to\infty}\sum_{\substack{k=1 \\ k\ne h_M}}^N
    \bigg| \frac{f(\frac kM)-f(\frac{h_M}{M})}{(k-h_M)^n} \bigg|=0, \qquad n\geq 2  \, .
 \end{equation}
\end{lemma}

\proof Let $n=2$. Then
 \begin{equation}\label{n=2}
 \sum_{\substack{k=1 \\ k\ne h_M}}^N\frac{|f(\frac kM)-f(\frac{h_M}{M})|}{(k-h_M)^2}
 \le \frac1{M^2}\sum_{\substack{k=1 \\ k\ne h_M}}^N
       \frac{|f(\frac kM)-f(\frac{h_M}{M})-\frac {k-h_M}{M}f'(\frac{h_M}{M})|}{(\frac{k-h_M}{M})^2}
       +
    \frac{|f'(\frac{h_M}{M})|}M\sum_{\substack{k=1 \\ k\ne h_M}}^N\frac1{k-h_M} \, .
 \end{equation}
The second term vanishes in the limit, as it is of order $\frac{\log
 N}M$. The first term vanishes due to the Euler--Maclaurin summation formula:
 \begin{multline}\label{n=2a}
 \lim_{ \substack{ N,M\to\infty \\ N/M \to D } }
 \frac1{M^2}\sum_{\substack{k=1 \\ k\ne h_M}}^N
       \frac{|f(\frac kM)-f(\frac{h_M}{M})-\frac {k-h_M}{M}f'(\frac{h_M}{M})|}{(\frac{k-h_M}{M})^2}
       \\
 %        \frac1{M^2} \sum_{k=1}^N\frac{f\left({\textstyle\frac    kM}\right)-f(0)-\frac kMf'(0)}{(\frac kM)^2}
 =\lim_{\substack{ N,M\to\infty \\ N/M \to D }}
    \frac1M\int\limits_0^D \! \frac{|f(x)-f(x_h)-(x-x_h)f'(x_h)|}{(x-x_h)^2}\,dx= 0.
 \end{multline}

For $n>2$, we have
\begin{equation}\label{n>2}
  \sum_{\substack{k=1 \\ k\ne h_M}}^N
    \bigg| \frac{f(\frac kM)-f(\frac{h_M}{M})}{(k-h_M)^n} \bigg|
  =\sum_{\substack{k=1 \\ k\ne h_M}}^N
     \bigg|\frac1{(k-h_M)^{n-2}}\cdot\frac{f(\frac kM)-f(\frac{h_M}{M})}{(k-h_M)^2} \bigg|
    \le
    \sum_{\substack{k=1 \\ k\ne h_M}}^N\frac{|f(\frac kM)-f(\frac{h_M}{M})|}{(k-h_M)^2}\to 0,
\end{equation}
which ends the proof.\qed

\begin{lemma}\label{sum-log-k-prep}
With the hypothesis of Lemma~\ref{m>1}, let $n_0\in\mathbb{N}$
be such that $\sup_{x\in[0,D]}|f(x)|<n_0$ and $|f(x_h)|<n_0$. Then
 \begin{equation}\label{log-kf}
S_{n_0;N}(f)
=\sum_{\substack{k=1 \\ |k-h_M| \ge n_0}}^N \log\frac{k-h_M+f(\frac kM)}{k-h_M+f(\frac{h_M}{M})}
        \underset{\substack{N,M \to \infty \\ \tf{N}{M}\to D} }{\longrightarrow}
   \int_0^D\frac{f(x)-f(x_h)}{x-x_h}\,dx \, .
 \end{equation}
\end{lemma}

\proof Expanding the logarithms into their Taylor series we obtain
 \begin{equation}\label{log-kfT}
S_{n_0;N}\pa{f}=\sum_{\substack{k=1 \\ |k-h_M| \ge n_0}}^N
    \sum_{r=1}^{\infty}\frac{(-1)^{r+1}}r\cdot \frac{f^r (\frac kM)-f^r(\frac{h_M}{M})}{(k-h_M)^r} .
    \end{equation}
The result of  Lemma~\ref{m>1} shows that only the terms
corresponding to $r=1$ give non-vanishing contributions. These are
computed as a Riemann sum:
 \begin{equation}\label{log-kfT-1}
\lim_{\substack{ N,M\to\infty \\ N/M \to D }
}S_{n_0;N}(f)=\lim_{\substack{ N,M\to\infty \\ N/M \to D } }
\frac1M\sum_{\substack{k=1 \\ |k-h_M| \ge n_0}}^N \frac{f(\frac kM)-f(\frac{h_M}{M})}{\frac {k-h_M}M}
 = \int_0^D\frac{f(x)-f(x_h)}{x-x_h}\,dx.
 \end{equation}
\qed

\begin{cor}\label{sing-product}
Let $f$ and $(h_M)$ satisfy the hypothesis of Lemma~\ref{m>1}.
Let $\xi(\la)$ be a strictly monotonous smooth function on $\R$
such that $\xi^{-1}(0)=-q$ and $\xi(D)=q$. Define
$\la_k=\xi^{-1}(k/M)$, $k \in \mathbb{Z}$ and
$\rho(\lambda) = \xi'(\la)$. Then, independently whether the
thermodynamic limit $\lambda_h$ of $\lambda_{h_M}$ belongs or not to the interval
$[-q,q]$, one has
 \begin{equation}\label{prod-sing}
 \lim_{\substack{ N,M\to\infty \\ N/M \to D }}\prod_{k=1}^N
 \frac{k-h_M+f(\lambda_k)}{k-h_M+f(\lambda_{h_M})}
 =
 \exp\left\{\int\limits_{-q}^q\frac{f(\lambda)-f(\lambda_h)}{\xi(\lambda)-\xi(\lambda_h)}
 \rho(\lambda)\,d\lambda\right\}.
 \end{equation}
\end{cor}

\proof

Assume that $h_M\in\{1,\dots,N\}$ and set $\wt{f}=f\circ\xi^{-1}$.
Choose $n_0> \sup_{x\in[0,D]}|\wt{f}(x)|$ and decompose the original
product into
 \begin{equation}\label{rep-prod}
 \prod_{k=1}^N\frac{k-h_M+f(\la_k) }{k-h_M+f(\la_h)}
 =\prod_{|k-h_M|<n_0}\frac{k-h_M+\wt{f}\big({\textstyle\frac kM}\big)}{k-h_M+\wt{f}\big(\frac {h_M}M\big)}
 \prod_{|k-h_M|\ge n_0}\frac{k-h_M+\wt{f}\big({\textstyle\frac kM}\big)}{k-h_M+\wt{f}\big(\frac {h_M}M\big)}.
\end{equation}
The first product contains a fixed finite number of factors. Each of
them goes to $1$ in the limit considered. Hence,  the first product
goes to $1$. The second can be calculated by using
Lemma~\ref{sum-log-k-prep}. We obtain
 \begin{equation}\label{prod-Fk}
 \lim_{\substack{ N,M\to\infty \\ N/M \to D }}\prod_{k=1}^N\frac{k-h_M+f(\la_k)}{k-h_M+f(\la_{h_M})}
=\exp\left\{\int_0^D\frac{\wt{f}(x)-\wt{f}(x_h)}{x-x_h}\,dx\right\},
 \end{equation}
where $x_h= \underset{N,M\to + \infty}{\lim} h_M/M$. Equation
\eqref{prod-sing} follows from \eqref{prod-Fk} after the change of
variable $\lambda=\xi^{-1}(x)$. The case of $h_M \not\in\paa{1,\dots,
N}$ is proven along the same lines. \qed

\section{The Fredholm determinant representations for the scalar
products\label{S-det-rep}}

The proof of \eqref{2-l-rep}, \eqref{Ps-B} is based on the following
representation for a scalar product \cite{Sla89,Sla95,KitMT99},
valid whenever $\mu_{\ell_1},\dots,\mu_{\ell_{N_{\kappa}}}$ satisfy
the system \eqref{TBE-cf} and $\nu_1,\dots,\nu_{N_{\kappa}}$ are
generic complex numbers:

\begin{equation}
\langle\psi_\kappa(\{\mu\})| \psi(\{\nu\})\rangle =
\frac{\prod\limits_{a=1}^{N_{\kappa}} d(\mu_{\ell_a})}
{\prod\limits_{a>b}^{N_{\kappa}}\sinh(\mu_{\ell_a}-\mu_{\ell_b})\sinh(\nu_b-\nu_a)}
\cdot \det_{N_{\kappa}}
\Omega_\kappa(\{\mu\},\{\nu\}|\{\mu\})\label{scal-prod} \, .
\end{equation}
The $N_{\kappa}\times N_{\kappa}$ matrix
$\Omega_\kappa(\{\mu\},\{\nu\}|\{\mu\})$  is defined as
\begin{multline} \label{matH}
  (\Omega_\kappa)_{jk}(\{\mu\},\{\nu\}|\{\mu\})=
  a(\nu_j)\,t(\mu_{\ell_k},\nu_j)\,\prod_{a=1}^{N_{\kappa}} \sinh(\mu_{\ell_a}-\nu_j-i\zeta)\\
   -\kappa\, d(\nu_j)\,t(\nu_j,\mu_{\ell_k})\,\prod_{a=1}^{N_{\kappa}} \sinh(\mu_{\ell_a}-\nu_j+i\zeta)\, ,
\end{multline}
with
\begin{equation}\label{def-t}
t(\mu,\nu)=\frac{-i\sin\zeta}{\sinh(\mu-\nu)\sinh(\mu-\nu-i\zeta)}
\; \quad \e{and}\quad
\left\{\begin{array}{c} a\pa{\mu}=\sinh^{M}\pa{\mu-i\tf{\zeta}{2}} \vspace{2mm}\\ d\pa{\mu}=\sinh^{M}\pa{\mu+i\tf{\zeta}{2}} \end{array}  \right.\;.
\end{equation}
In order to obtain the scalar product \eqref{2-l-rep} one should set
here $\nu_j=\lambda_j$ for $j=1,\dots,N$, where $\lambda_j$ are the
Bethe roots describing the ground state. For the scalar product
\eqref{Ps-B} one should set in addition $\nu_{N+1}=-i\zeta/2$.

To obtain a Fredholm determinant representation for the scalar
products we should present the original determinant of the matrix
$\Omega_\kappa$ in the following form
\begin{equation}\label{Transf-SP}
 \det_{N_{\kappa}}
 \bigl.\Omega_\kappa(\{\mu\},\{\nu\}|\{\mu\})\bigr|_{\nu_j=\lambda_j}=H(\{\mu\},\{\lambda\})\det_N\Bigl(\delta_{jk}+
 \widetilde\Omega(\lambda_j,\lambda_k|\{\mu\})\prod_{a=1,\atop{a\ne j}}^N(\lambda_j-\lambda_a)^{-1}\Bigr).
\end{equation}
Here $\widetilde\Omega(\lambda_j,\lambda_k|\{\mu\})$ is a new
$N\times N$ matrix, $H(\{\mu\},\{\lambda\})$ an external
coefficient. If we succeed to find a representation of the type
\eqref{Transf-SP}, then we can replace the  determinant of the
$N\times N$ matrix by Fredholm determinant of an integral operator
as
\begin{equation}\label{SP-FD}
 \det_N\Bigl(\delta_{jk}+
 \widetilde\Omega(\lambda_j,\lambda_k|\{\mu\})\prod_{a=1,\atop{a\ne
 j}}^N(\lambda_j-\lambda_a)^{-1}\Bigr)=
 \det_{\Gamma_q}\Bigl(I+\frac1{2\pi i}
 \widetilde\Omega(w,w'|\{\mu\})\prod_{a=1}^N(w-\lambda_a)^{-1}\Bigr).
\end{equation}
Here the  integral operator in the r.h.s. of \eqref{SP-FD} acts on a
counterclockwise oriented contour $\Gamma_q$ surrounding the Fermi
zone $[-q,q]$. This contour is such that it contains all the ground
state roots $\lambda_j$ and no other singularity of the kernel
$\widetilde\Omega(w,w'|\{\mu\})$.

The proof of \eqref{SP-FD} is quite obvious. Indeed, expanding the
Fredholm determinant into the series of multiple integrals we see
that each of these integrals reduces to the sum of the residues in
the points $\lambda_j$. Thus, the series of multiple integrals turns
into the series of multiple sums. Then one can easily convince
oneself that this series coincides with the expansion of the
$N\times N$ determinant in the l.h.s. of \eqref{SP-FD}.

Thus, our goal is to pass from  the original representation
\eqref{scal-prod} to the form \eqref{Transf-SP}. The way we use here
is based on the extraction of a Cauchy determinant from the original
determinant defining the scalar product, an idea which was first
introduced in \cite{Sla90,IzeKMT99}. In the case of the scalar
product \eqref{2-l-rep} this was done in \cite{KitKMST09b}. The same
method with minor modifications described below can be used for the
proof of the second determinant identity \eqref{Ps-B}.

Following \cite{KitKMST09b}, we consider the matrix
$\Om_{\kappa}(\{z\},\{\nu\}|\{z\})$ for  two sets of
$N_{\kappa}=N+1$ generic parameters $\{z\}$ and $\{\nu\}$. In this
case we write
\begin{equation}
\det_{N+1}\pac{ \Om_{\kappa}  }= \frac{\det_{N+1}(\Om_{\kappa} A)}{\det_{N+1} A} \quad \text{with}\
 A_{jk}=\frac{\prod\limits_{a=1}^{N+1}\sinh(z_j-\nu_a)}
 {\prod\limits_{\substack{a=1\\ a\ne j}}^{N+1}\sinh(z_j-z_a)}\times
 \begin{cases}
 \coth(z_j-\nu_k) &\text{for}\ k\le N, \\
 1 &\text{for}\ k=N+1.
 \end{cases}
 %\left\{ \begin{array}{ll} \!\! \coth(z_j-\nu_k)\ &\text{for}\ k\ne p\num  \!\! 1\ &\text{for}\ k=p  \end{array}\right.   .
 \end{equation}
The effect of multiplication by the matrix $A$ is computed similarly to the method presented in \cite{KitKMST09b}. We get
\begin{equation}
\det_{N+1}\pac{\Om_{\kappa}(\{z\},\{\nu\}|\{z\}) }= \det_{N+1}\pac{ \f{1}{\s{z_k-\nu_j}}} \cdot \det_{N+1}{S_{jk}}
        \;,
\end{equation}
where
\begin{equation}\label{Sjk}
 S_{jk}=\delta_{jk}{\cal Y}_\kappa(\nu_j|\{z\})+\frac{%
 \prod_{a=1}^{N+1}\sinh(\nu_k-z_a)}{\prod_{a=1\atop{a\ne k}}^{N+1}
 \sinh(\nu_k-\nu_a)}\cdot\left.\frac\partial{\partial y_k}
 {\cal Y}_\kappa(\nu_j|\{ y \})\right|_{\paa{y}=\paa{\nu}}, \qquad
 k\le N \; ,
\end{equation}
\begin{equation}\label{Spk}
 S_{j,N+1}={\cal Y}_\kappa(\nu_j|\{\nu\}) \; ,
\end{equation}
and we have set for arbitrary complex $y_1,\dots,y_{N+1}$
\begin{equation}\label{TTM_Y-def}
   {\cal Y}_\kappa(\nu|\{y\}) =
      a(\nu)\prod_{k=1}^{N+1}\sinh(y_k-\nu-i\zeta)
      + \kappa\,d(\nu) \prod_{k=1}^{N+1}\sinh(y_k-\nu+i\zeta) \;.
\end{equation}
Finally we reduce the size of $\det_{N+1}S$ by one by performing
linear combinations of the lines
\begin{equation}
\det_{N+1}\pac{S_{jk}}=S_{N+1,N+1}\det_{N}\left[S_{jk}-S_{j,N+1}
\frac{S_{N+1,k}}{S_{N+1,N+1}} \right] \;.
\end{equation}
Thus, we arrive at the following representation for
$\det_{N+1}\Om_{\kappa}$ depending on generic complex
$z_1,\dots,z_{N+1}$ and $\nu_1,\dots,\nu_{N+1}$:
\begin{equation}\label{Om-newrep}
\det_{N+1}\pac{\Om_{\kappa}(\{z\},\{\nu\}|\{z\}) }=
S_{N+1,N+1}\det_{N+1}\pac{
\f{1}{\s{z_k-\nu_j}}}\det_{N}\left[S_{jk}-S_{j,N+1}
\frac{S_{N+1,k}}{S_{N+1,N+1}} \right] \;,
\end{equation}
where $S_{jk}$ are given by \eqref{Sjk}, \eqref{Spk}.

Now we set $z_a=\{\mu_{\ell_a}\}$ for $a=1,\dots, N+1$,
$\nu_k=\la_k$ for $k=1,\dots,N$  and $\nu_{N+1}=-i\tf{\zeta}{2}$. We
assume that the parameters $\{\mu_{\ell_a}\}$ satisfy the system
\eqref{TBE-lj}, while the parameters $\{\lambda_{k}\}$ satisfy the
system \eqref{BE}. Then we obtain for $j\le N$,
\begin{equation}\label{nu-la-j}
\begin{array}{l}
{\dis \frac{\mc{Y}_{\kappa}\pa{\nu_j\mid \{\nu\}}
}{a(\lambda_j)\prod_{b=1}^N\sinh(\lambda_b-\lambda_j-i\zeta)}=\kappa
\sinh(\lambda_j-{\textstyle\frac{i\zeta}2})
-\sinh(\lambda_j+{\textstyle\frac{3i\zeta}2}),}\num
{\dis \left.\frac{\partial\mc{Y}_{\kappa}\pa{\nu_j\mid \{y
\}}/\partial{y_k}
}{a(\lambda_j)\prod_{b=1}^N\sinh(\lambda_b-\lambda_j-i\zeta)}\right|_{\{y\}=\{\nu\}}=\kappa
 \frac{\sinh(\lambda_j-{\textstyle\frac{i\zeta}2})}{\tanh(\lambda_k-\lambda_j+i\zeta)}
-\frac{\sinh(\lambda_j+{\textstyle\frac{3i\zeta}2})}{\tanh(\lambda_k-\lambda_j-i\zeta)}},
  \end{array}
\end{equation}
and for $j= N+1$,
\begin{equation}\label{nu-la-N+1}
\begin{array}{l}
{\dis \mc{Y}_{\kappa}\pa{\nu_{N+1}\mid \{\nu\}}
 =a(-i\zeta/2)\sinh(-i\zeta)\prod_{b=1}^N\sinh(\lambda_b-{\textstyle\frac{i\zeta}2}),}\num
{\dis \Bigl.\partial\mc{Y}_{\kappa}\pa{\nu_{N+1}\mid \{y
\}}/\partial{y_k} \Bigr|_{\{y\}=\{\nu\}}=
 a(-i\zeta/2)\sinh(-i\zeta)\coth(\lambda_k-{\textstyle\frac{i\zeta}2})\prod_{b=1}^N\sinh(\lambda_b-{\textstyle\frac{i\zeta}2})}.
  \end{array}
\end{equation}
It also follows from \eqref{BE-cf}, \eq{TBE-cf} and the definition
of the shift function \eqref{SP-shiftF-mod} that
\begin{equation}
 \kappa\prod_{a=1}^{N+1}\frac{\sinh(\mu_{\ell_a}-w+i\zeta)}{\sinh(\mu_{\ell_a}-w-i\zeta)}
  \prod_{a=1}^{N}\frac{\sinh(\lambda_{a}-w-i\zeta)}{\sinh(\lambda_{a}-w+i\zeta)}=
  e^{2\pi i\widehat{F}(w)}\;.
\end{equation}
Substituting all these formulae into \eqref{Om-newrep} we arrive
after simple algebra at the representation of the form
\eqref{Transf-SP}, and thus, we obtain the Fredholm determinant
representation \eqref{Ps-B}. \qed

\end{document}